\documentclass[journal,10pt]{IEEEtran}

\usepackage{amsmath,amssymb}
\usepackage{subfigure}
\usepackage{graphicx,graphics,color,psfrag}
\usepackage{cite,balance}
\usepackage{algorithm}
\usepackage{accents}
\usepackage{amsthm}
\usepackage{bm}
\usepackage{url}
\usepackage{algorithmic}
\usepackage[english]{babel}
\addto\captionsenglish{}
\usepackage{multirow}
\usepackage{enumerate}
\usepackage{cases}
\usepackage{stfloats}
\usepackage{placeins}
\usepackage{dsfont}
\usepackage{color,soul}
\usepackage{amsfonts}
\usepackage{tcolorbox}

\usepackage{booktabs} 
\usepackage{array}
\usepackage{tabularx} 
\usepackage{algorithm}
\usepackage{acronym} 
\usepackage{makecell}

\usepackage{cite,graphicx,amsmath,amssymb}
\usepackage{fancyhdr}
\usepackage{hhline}
\usepackage{graphicx,graphics}
\usepackage{array,color}
\usepackage{amsmath}
\usepackage{amsthm}
\usepackage[flushleft]{threeparttable}
\IEEEoverridecommandlockouts

\newtheoremstyle{propositionstyle}
{\topsep}
{\topsep}
{\normalfont}
{}
{\normalfont}
{.}
{0.5em}
{\thmname{\textit{#1}}\thmnumber{\textit{ #2}}\thmnote{\normalfont\ (#3)}}
\theoremstyle{propositionstyle}
\newtheorem{proposition}{Proposition}

\theoremstyle{plain}

\makeatletter
\def\endthebibliography{%
	\def\@noitemerr{\@latex@warning{Empty `thebibliography' environment}}%
	\endlist
}
\makeatother

\newif\ifcommapvbenchmark
\commapvbenchmarktrue

\begin{document}
\title{Movable Antenna\textcolor{black}{--}Aided ISAC: A Time\textcolor{black}{--} and Frequency\textcolor{black}{--}Division \textcolor{black}{Framework}}
\author{Yuyan Zhou, \emph{Student Member, IEEE,} \,\thanks{Y. Zhou is with the School of Science and Engineering, The Chinese University of Hong Kong (Shenzhen), Guangdong 518172, China (email: yuyanzhou1@link.cuhk.edu.cn).} Haocheng Hua, \emph{Member, IEEE,} \thanks{H. Hua is with the College of Electronics and Information Engineering,
Shenzhen University, Shenzhen, China (email: huahaocheng@szu.edu.cn).} Jie Xu, \emph{Fellow, IEEE,} \thanks{J. Xu is with the School of Science and Engineering, the Shenzhen Future Network of Intelligence Institute (FNii-Shenzhen), and the Guangdong Provincial Key Laboratory of Future Networks of Intelligence, The Chinese University of Hong Kong (Shenzhen), Guangdong 518172, China
(email: xujie@cuhk.edu.cn).} and Rui Zhang, \emph{Fellow, IEEE} \thanks{R. Zhang is with the Department of Electrical and Computer Engineering, National University of Singapore, Singapore 117583 (e-mail: elezhang@nus.edu.sg).} \thanks{Haocheng Hua is the corresponding author.}}

\maketitle
\begin{abstract}
This paper investigates movable-antenna (MA)-enhanced integrated sensing and communication (ISAC) systems under time division duplex (TDD) and frequency division duplex (FDD) protocols, in which sensing and communications are implemented over orthogonal time and frequency resources, respectively. For sensing, a wideband performance metric is established by deriving the target-position Cram\'er-Rao bound (CRB) under a pulse-radar framework with pulse compression. The resulting position CRB incorporates delay and angle estimation error bounds in a Cartesian coordinate system. Both error bounds depend on sensing power, while \textcolor{black}{in addition}, the delay portion depends on sensing bandwidth and the angle portion depends on the effective aperture determined by the antenna position vector (APV). A worst-case position CRB over a prescribed target region is adopted to ensure robust sensing performance. For communication, a multiuser downlink scenario is considered, where the minimum achievable rate among all users captures user fairness. Based on these metrics, TDD- \textcolor{black}{or} FDD-based joint APV and resource-allocation problems are formulated to maximize the \textcolor{black}{worst-case} common communication rate \textcolor{black}{across users} subject to the \textcolor{black}{given} sensing requirement. An alternating optimization algorithm is developed for the FDD formulation, while the TDD formulation is solved through a search over all the feasible numbers of sensing pulses combined with APV optimization. 
{Moreover, representative scenarios are analyzed to clarify the roles of MA in improving channel orthogonality and enlarging the effective aperture for wireless sensing, as well as the resulting rate comparison under TDD/FDD schemes. Numerical results show that APV optimization enlarges the communication-sensing tradeoff region over fixed-position antenna benchmarks and reveals distinct effects of APV optimization between the two schemes, with TDD tending to favor a larger antenna position variance and FDD tending to favor better channel orthogonality among different users.}
\end{abstract}

\section{Introduction}
The next generation of wireless systems is expected to provide not only high data rates and reliable communications, but also environment-aware services, which require wireless networks to possess advanced sensing capabilities \cite{6G_survey}. This has stimulated growing interest in integrated sensing and communication (ISAC), which enables communication and sensing functionalities to be realized on a common wireless platform by sharing hardware and radio resources. Owing to its capability of unifying these two functions within the same infrastructure, ISAC has been widely recognized as a key enabling technology for future 6G systems \cite{ISAC_overview}.

Most existing ISAC studies focus on highly integrated ISAC designs, where communication and sensing are tightly coupled in hardware, spectrum, and waveform design. Transmit beamforming, transmit covariance optimization, and waveform design have been studied to balance communication rate and sensing accuracy \cite{ISAC_1}, \cite{ISAC_2}, \cite{CRB_beamforming}. Such tight integration, however, mixes communication signals and sensing echoes and might incur \textcolor{black}{severe} self-interference in the practical deployment. Alternatively, resource-orthogonalized ISAC separates the two functions in time or frequency: time division duplex (TDD) ISAC schedules them in distinct intervals \cite{joint_design_comm_sensing_6g}, whereas frequency division duplex (FDD) ISAC assigns them to different subcarriers or frequency bands \cite{ofdm_radar_comm}, \cite{11}. Although separation mitigates mutual interference, it introduces a resource tradeoff. Moreover, most existing studies rely on fixed-position antenna (FPA) arrays, whose fixed geometries cannot simultaneously adapt to the spatial requirements of multiuser communication and target sensing. 
While communication benefits from low inter-user channel correlation with reduced cross-channel interference, sensing requires a large effective aperture and unambiguous spatial responses.
This motivates antenna position adaptation as an additional degree of freedom (DoF) for ISAC.

Movable antenna (MA) \cite{MA_overview} has recently emerged as a promising technology for wireless systems by introducing antenna position as an additional design DoF. Different from conventional FPA arrays, MA-enabled transceivers can adjust the antenna locations within a given region to exploit the spatial variation of wireless channels. This capability has been studied in various wireless communication scenarios \cite{MA_modeling}, \cite{MA_interference_network}. Beyond position optimization, rotatable antenna (RA) \cite{RA_Zheng} and six-dimensional MA (6DMA)\textcolor{black}{\cite{6DMA1,6DMA,6DMA2}} systems introduce antenna rotation as an additional design dimension, further enlarging the spatial design space.

The benefits of MA have also been investigated in wireless sensing systems. By flexibly adjusting antenna positions, MA arrays can synthesize a larger effective aperture, thereby improving angular resolution \cite{MA_sensing}. Furthermore, antenna trajectory optimization has been studied to characterize the Cram\'er-Rao bound (CRB) scaling behavior over time and space \cite{MA_sensing_traj}. 
These studies show that MA can improve sensing performance by optimizing the array geometry, without necessarily increasing the number of antenna elements or radio-frequency (RF) chains.

The above progress has further motivated research on MA/6DMA-enabled ISAC systems. 
For communication-oriented designs, joint antenna position and beamforming optimization has been investigated to improve communication rates under target angle estimation CRB constraints \cite{MA_ISAC_Ma} and a sensing signal-to-interference-plus-noise ratio (SINR) constraint \cite{MA_ISAC_networked_rate}, \cite{MA_ISAC_secure_RIS}. 
For sensing-oriented designs, joint beamforming and MA position optimization has been employed to minimize the target angle estimation CRB under communication SINR constraints \cite{MA_ISAC_Qin}.
These works demonstrate that MA can improve the communication-sensing tradeoff by introducing geometry-domain optimization beyond conventional beamforming and power control.

However, most existing MA-enabled ISAC studies adopt narrowband sensing models and thus do not explicitly account for the impact of sensing bandwidth on delay estimation. This limitation is important for wideband localization \cite{loca}. Under resource-orthogonalized ISAC, TDD and FDD introduce fundamentally different sensing-resource tradeoffs: TDD preserves the full sensing bandwidth and transmit power at the expense of communication time slots, whereas FDD enables sensing-communication time overlap by partitioning the available bandwidth and power. How to jointly optimize time, bandwidth, power, and antenna positions while characterizing the performance difference between the two schemes remains largely unexplored.

Motivated by this gap, this paper investigates an MA-aided ISAC system with wideband target sensing and TDD-/FDD-based resource orthogonalization. The antenna positions and scheme-specific resource allocation are jointly optimized to maximize the \textcolor{black}{worst-case} communication rate \textcolor{black}{of users} subject to a robust target-position CRB constraint \textcolor{black}{for sensing}. The main contributions are summarized as follows.
\begin{itemize}
\item {First, we establish a wideband performance model for MA-aided ISAC. A target-position CRB is derived under a pulse-radar model with pulse compression and transformed into a Cartesian coordinate system, accounting for both delay- and angle-estimation error bounds. Both error bounds depend on sensing power. The delay component additionally depends on sensing bandwidth, while the angle component \textcolor{black}{critically} depends on the aperture induced by antenna positions. A worst-case CRB over a target uncertainty region is then obtained as a tractable robust sensing metric.}
\item {Next, we formulate both TDD- and FDD-based optimization problems for the joint antenna position optimization and resource allocation. The FDD-based formulation is solved by alternating optimization, where bandwidth/power allocation is obtained from the Karush-Kuhn-Tucker (KKT) conditions and the antenna position is updated by projected gradient with Dykstra projection. The TDD-based formulation is addressed by an outer search over the sensing pulse number combined with \textcolor{black}{the} inner antenna position optimization.}
\item Furthermore, representative scenarios with different user angular separations and relative contributions of delay- and angle-related portions to the sensing CRB are analyzed to reveal how MA improves channel orthogonality and enlarges the sensing aperture. TDD generally performs better when the delay-related portion dominates, whereas FDD tends to be more favorable when the angle-related portion dominates. However, FDD bandwidth partitioning increases the relative contribution of the delay-related portion, potentially weakening its advantage.
\item Finally, extensive numerical results demonstrate that antenna position optimization enlarges the communication-sensing tradeoff region over FPA benchmarks. Increasing the total bandwidth or the target range tends to favor FDD over TDD in terms of achievable communication rate. The optimized antenna position reconfigurations exhibit distinct characteristics, with TDD tending to favor a larger antenna position variance and FDD tending to favor better channel orthogonality among different users.
\end{itemize}

{The rest of this paper is organized as follows. Section II presents the system model and the problem formulations. Sections III and IV develop the FDD- and TDD-based algorithms, respectively. Section V compares the two schemes, Section VI presents numerical results, and Section VII concludes this paper.}

\textit{Notations:} Boldface lower-case and upper-case letters denote vectors and matrices, respectively. For a matrix $\bm M$, $\bm M^T$, $\bm M^H$, and $\bm M^{-1}$ denote its transpose, conjugate transpose, and inverse. $\mathbb{R}^{x\times y}$ and $\mathbb{C}^{x\times y}$ denote real and complex matrix spaces, respectively. $\mathbb{E}\{\cdot\}$, $\mathrm{tr}(\cdot)$, $\mathrm{Re}\{\cdot\}$, and $\mathrm{Im}\{\cdot\}$ denote expectation, trace, real part, and imaginary part, respectively. $\|\cdot\|$, $\|\cdot\|_F$, and $\odot$ denote the Euclidean norm, Frobenius norm, and Hadamard product. $\bm I_N$ and $\bm 1_N$ denote the $N\times N$ identity matrix and the $N$-dimensional all-one vector. $\mathcal{CN}(0,\sigma^2)$ denotes the circularly symmetric complex Gaussian distribution with zero mean and variance $\sigma^2$, $\Pi_{\mathcal{C}}(\cdot)$ denotes projection onto set $\mathcal C$ and $j=\sqrt{-1}$. $\lfloor x\rfloor$ denotes the floor function, i.e., the greatest integer less than or equal to $x$.

\section{System Model}
\label{Section_system_model}
\begin{figure}[t]
\centering
\includegraphics[width=3.5in, height = 1.6 in]{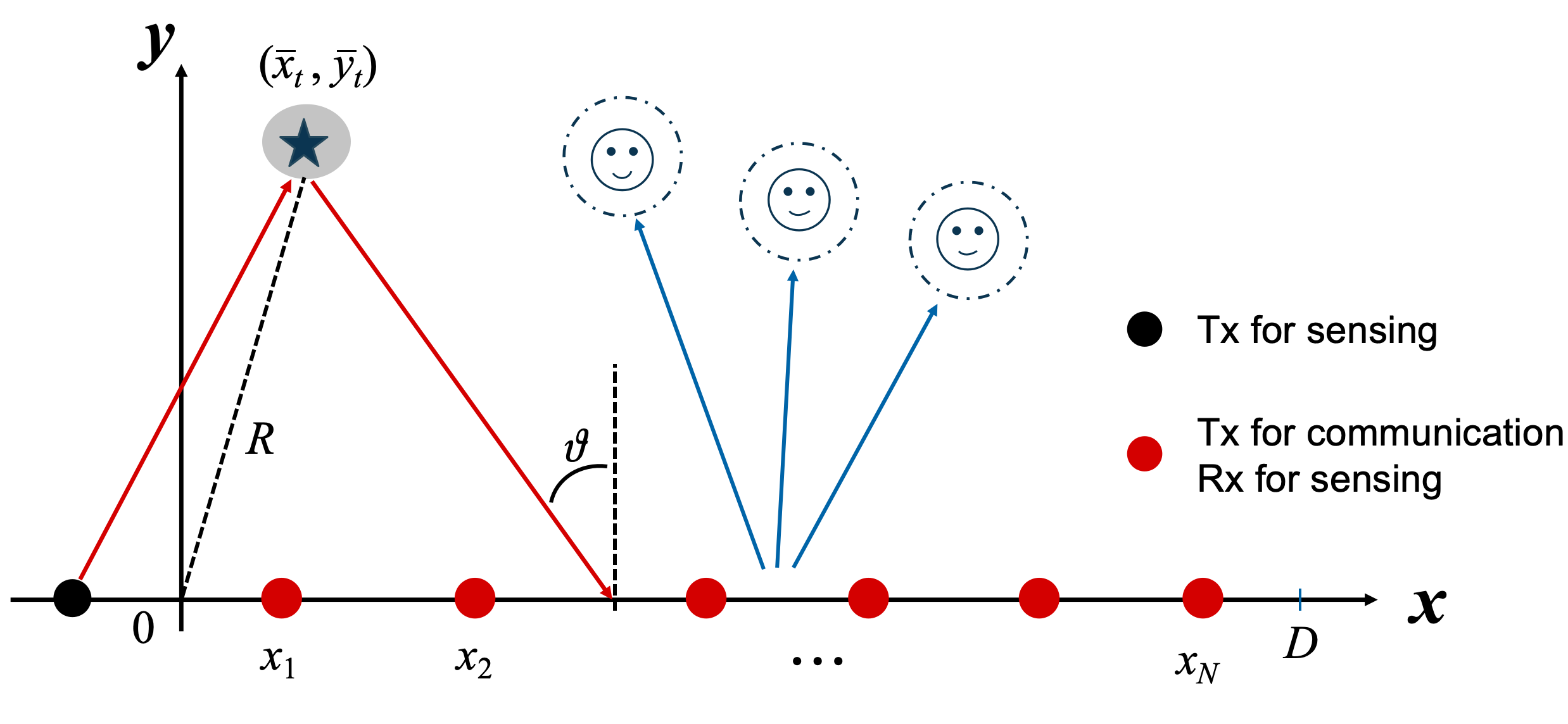}
\vspace{-0.4cm}
\caption{The illustration of an ISAC base station equipped with MA.}
\label{fig:system}
\vspace{0.3cm}
\end{figure}

Fig.~\ref{fig:system} illustrates the considered MA-aided ISAC system, where the base station (BS) employs an $N$-element MA array to serve $K$ single-antenna downlink users and to collect the echo reflected by one sensing target. We assume that the number of users is smaller than the number of MAs, i.e., $K<N$. A separate fixed antenna is used to transmit the sensing signal to illuminate the target, whereas the MA array is shared by the downlink transmitter and the sensing receiver. In general, communication and sensing may be separated in either the time domain or the frequency domain. To establish a unified formulation, let $T_c$, $B_c$, and $P_c$ denote the transmission time, bandwidth, and transmit power assigned to communication, respectively. Similarly, let $T_s$, $B_s$, and $P_s$ denote the transmission time, bandwidth, and transmit power assigned to sensing, respectively. The total transmission frame duration, total available bandwidth, and total transmit power are denoted by $T_{\rm total}$, $B_T$, and $P_T$, respectively. The total frame is further quantized into $Q$ short time slots, each with duration $T_{\rm slot}$, such that $T_{\rm total}=Q T_{\rm slot}$.

The MA elements are constrained to a one-dimensional region $[0,D]$. Their locations are collected in the antenna position vector (APV) $\bm{x}=[x_1,\ldots,x_i,\ldots,x_N]^T$, where $x_i$ denotes the coordinate of the $i$-th MA. Based on this geometry, the downlink communication and target sensing metrics are developed in the following.

\vspace{0.05cm}
\subsection{Multiuser Downlink Communication}
We first present the downlink communication metric over the communication resource block specified by $(T_c,B_c,P_c)$. Under the slotted frame model, let $N_c$ denote the number of time slots assigned to communication, such that $T_c=N_cT_{\rm slot}$ and $T_c/T_{\rm total}=N_c/Q$. Let $\mathcal K\triangleq\{1,\ldots,K\}$ denote the user set. The BS transmits one data stream to each user, where $\bm w_k\in\mathbb C^{N\times 1}$ is the beamforming vector for the $k$-th user and $s_k\in\mathbb C$ is the corresponding information-bearing symbol. The symbols are mutually independent, zero mean and normalized to unit average power, i.e., $\mathbb{E}\{s_k\}=0 $, $\mathbb E\{|s_k|^2\}=1$ and $\mathbb E\{s_k s_j^*\}=0$ for $k\ne j$. The communication signal transmitted by the MA array is given as
\begin{equation}
\bm{s}_c = {\sum}_{k=1}^K \bm{w}_k s_k,
\end{equation}
where the beamforming/\textcolor{black}{precoding} matrix $\bm W=[\bm w_1,\ldots,\bm w_K]\in\mathbb C^{N\times K}$ satisfies the communication power budget 
\begin{equation}
\|\bm W\|_F^2\le P_c.
\end{equation}

For the $k$-th user, the BS-user link is modeled as a line-of-sight (LoS) channel\footnote{Since ISAC systems are typically operated in high frequency bands, such as millimeter-wave bands, the BS-user channel is often LoS-dominated. The proposed \textcolor{black}{communication system model} can also be extended to a general multi-path channel model.}. The channel vector is expressed as
\begin{align}
\bm{h}_k(\bm{x},\theta_k)
&= \beta_k\left[e^{j \frac{2\pi}{\lambda} x_1 \sin\theta_k},\cdots,e^{j \frac{2\pi}{\lambda} x_N \sin\theta_k}\right]^T \\
&=\beta_k \bm{a}(\bm{x},\theta_k),
\end{align}
where $\bm a(\bm x,\theta_k)$ is the MA array steering vector, $\beta_k=\frac{\lambda}{4\pi d_k}e^{-j\frac{2\pi}{\lambda}d_k}$ is the complex path gain, \(\lambda\) denotes the carrier wavelength, and $\theta_k$ and $d_k$ denote the angle and distance of user $k$ with respect to the original point of the antenna array equipped at the BS, respectively.
The baseband signal received by the $k$-th user is expressed as
\begin{equation}
y_k=\bm{h}_k^H(\bm{x},\theta_k)\sum_{i=1}^K \bm{w}_i s_i+n_k,
\label{equ:Rx_signal}
\end{equation}
where $n_k\sim\mathcal{CN}(0,N_0B_c)$ is the additive white Gaussian noise and $N_0$ is the noise power spectral density in the communication subsystem. Consequently, the SINR for the $k$-th user is given by
\begin{equation}
{\rm{SINR}}_k(\bm{x},\bm{W},B_c)
=
\frac{|\bm{w}_k^H \bm{h}_k(\bm{x},\theta_k)|^2}
{\sum_{i\ne k} |\bm{w}_i^H \bm{h}_k(\bm{x},\theta_k)|^2 + N_0 B_c}.
\end{equation}
The instantaneous achievable rate during the communication interval $T_{c}$ is
\begin{equation}
\tilde R_k(\bm{x},\bm W,B_c)
=
B_c\log_2\big(1+{\rm{SINR}}_k(\bm{x},\bm W,B_c)\big).
\end{equation}
Accordingly, the corresponding average rate of the $k$-th user over the total frame ${T_{\rm total}}$ is
\begin{equation}
R_k(\bm{x},\bm W,N_c,B_c)
=
\frac{N_c}{Q}B_c
\log_2\big(1+{\rm{SINR}}_k(\bm{x},\bm W,B_c)\big).
\end{equation}
Since the system serves multiple users, the \textcolor{black}{worst-case} common frame-average rate is adopted as \textcolor{black}{a} user-fairness metric, which is given by
\begin{equation}
f^{\rm orig}_c(\bm W,N_c,B_c,\bm x)
=
\min_{k\in\mathcal K} R_k(\bm{x},\bm W,N_c,B_c).
\label{eq:min-rate-general}
\end{equation}

To avoid the highly coupled joint optimization of beamforming, antenna positions, and resource allocation, zero-forcing (ZF) precoding is adopted for multiuser \textcolor{black}{downlink} transmission. The normalized ZF precoder is given by
\begin{equation}
\bm W_{\mathrm{ZF}}
=
\sqrt{P_c}\,
\frac{\bm H(\bm x,\bm\theta)
\big(\bm H^H(\bm x,\bm\theta)\bm H(\bm x,\bm\theta)\big)^{-1}}
{\left\|\bm H(\bm x,\bm\theta)
\big(\bm H^H(\bm x,\bm\theta)\bm H(\bm x,\bm\theta)\big)^{-1}\right\|_F},
\end{equation}
where $\bm H(\bm x,\bm\theta)=[\bm {h}_1,\ldots,\bm{h}_K]$ denotes the channel matrix between the BS and all users, and $\bm\theta=[\theta_1,\ldots,\theta_K]^T$ denotes the angle-of-departure (AoD) of the users. With this normalization, the total communication power satisfies $\|\bm W_{\mathrm{ZF}}\|_F^2=P_c$, and all users achieve the same signal-to-noise ratio (SNR), which is given by
\begin{equation}
\tilde{\gamma}(\bm x,P_c,B_c)
=
\frac{P_c}
{\mathrm{tr}\!\left[
\big(\bm H^H(\bm x,\bm\theta)\bm H(\bm x,\bm\theta)\big)^{-1}
\right] N_0 B_c}.
\label{eq:gamma-general}
\end{equation}
Accordingly, the ZF-based common frame-average rate is expressed as
\begin{equation}
f^{\rm ZF}_c(\bm x,N_c,B_c,P_c)
=
\frac{N_c}{Q} B_c
\log_2\left(1+\tilde{\gamma}(\bm x,P_c,B_c)\right).
\label{eq:fc-zf-general}
\end{equation}

\begin{figure*}[t]
\centering
\includegraphics[width=6.57in]{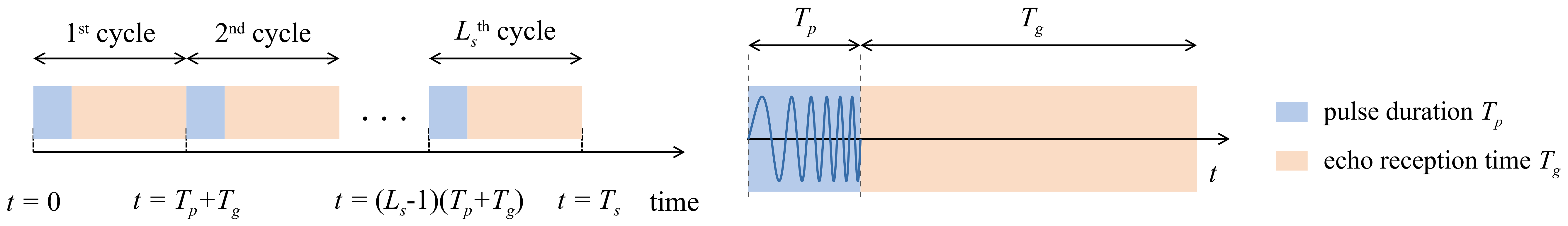}
\caption{Multiple sensing pulse timeline within the sensing frame $T_s$ and linear frequency modulation waveform illustration.}
\label{fig:sensing_pulse_timeline}
\end{figure*}

\subsection{Target Sensing}
For sensing, the target position is first described by the polar-domain parameters $(R,\vartheta)$, where $R$ denotes the range from the BS and $\vartheta$ denotes the angle of arrival (AoA) measured from the array broadside. The target range is assumed to be much larger than the array aperture, so $R$ and $\vartheta$ remain approximately unchanged across the array. Since the fixed illuminating antenna is co-located with the MA receiving array, the round-trip propagation delay is approximated as $\tau=2R/c$, where $c$ is the speed of light. Let $s_r(t)$ denote the sensing waveform with pulse duration $T_p$ and bandwidth $B_s$. Its average power is normalized to unity, i.e., $\frac{1}{T_p}\int_{0}^{T_p}|s_r(t)|^2dt=1$, while the actual sensing signal is transmitted with power $P_s$.

Within the sensing duration $T_s$, multiple sensing pulses are transmitted for coherent processing, as shown in Fig. 2. Each sensing-pulse cycle consists of pulse transmission and echo reception, and its pulse repetition interval (PRI) is
\begin{equation}
T_{\rm PRI}=T_p+T_g,
\label{eq:PRI-general}
\end{equation}
where $T_g$ is the echo-reception interval. For simplicity, we assume that one sensing pulse cycle occupies $L = \frac{T_{\rm PRI}}{T_{\rm slot}} \in\mathbb Z_{+}$
time slots. In addition, the sensing interval accommodates a total of $L_s$ pulses.

The echo corresponding to the $\ell$-th sensing pulse can be written as
\begin{align}
\bm{y}_{\ell}(t)
&=
\alpha \sqrt{P_s} \bm{a}(\bm{x}, \vartheta)
s_r(t-t_\ell-\tau)
+ \bm{z}_{\ell}(t),
\end{align}
where {$t_\ell=(\ell-1)LT_{\rm slot}$} is the starting time of the $\ell$-th pulse, $\alpha$ is the complex target reflection coefficient including propagation loss, and $\bm z_\ell(t)$ denotes the additive noise vector.

For the single sensing pulse case, the receiver estimates the delay by applying matched filtering to the known sensing waveform and combining the echo over the MA array. The corresponding delay-estimation CRB is characterized by \cite{kay1993}
\begin{equation}
\mathrm{CRB}_{\tau}^{(1)}
=
\frac{1}
{4\pi^2 {\tilde{\rm{SNR}}} N\beta^2}.
\label{eq:CRB-tau-general}
\end{equation}
In \eqref{eq:CRB-tau-general}, the matched-filter output SNR is given as
\begin{equation}
{\tilde{\rm{SNR}}} = \frac{|\alpha|^2 P_s T_p}{N_0/2}, \label{se.snr}
\end{equation}
where $N_0$ denotes the noise power spectral density, which is assumed to be the same as that in the communication subsystem.
The squared effective bandwidth is given by
\begin{equation}
\beta^2 = \frac{\int_{-\infty}^{\infty} f^2 |S_r(f)|^2 df}{\int_{-\infty}^{\infty} |S_r(f)|^2 df} - \frac{(\int_{-\infty}^{\infty} f |S_r(f)|^2 df )^2} {(\int_{-\infty}^{\infty} |S_r(f)|^2 df)^2},
\end{equation}
where $S_r(f)=\int_{-\infty}^{\infty}s_r(t)e^{-j2\pi ft}dt$ is the Fourier transform of the transmitted sensing waveform.

{For a plain rectangular pulse, the delay/range resolution is mainly limited by the pulse width $T_p$, i.e.,
\begin{equation}
\Delta R \approx \frac{c T_p}{2}\textcolor{black}{,}
\label{eq:range-resolution}
\end{equation}
which reveals the fundamental tradeoff between echo SNR and range resolution: increasing the pulse width $T_p$ improves the receiving echo SNR but worsens the resolution limit, and vice versa. To avoid this limitation, we adopt pulse compression. In particular, linear frequency modulation (LFM) is a commonly adopted pulse-compression technique, where the pulse is modulated over a bandwidth of $B_s$ within the pulse width $T_p$, as shown in Fig. 2.}
The LFM waveform can be written as
\begin{equation}
s_r(t) = A \cos\left(2\pi f_{\mathrm{start}} t + \pi \mu t^2\right), \quad 0 \leq t \leq T_p,
\end{equation}
where  $f_{\mathrm{start}} = f_0 - B_s/2$ is the starting frequency, $f_0 = c/\lambda$ is the center frequency, $\mu = B_s/T_p$ is the chirp rate, and $A$ normalizes the pulse power.
After matched filtering, the energy of this long-duration chirp is compressed into a narrow correlation peak and the range resolution is mainly determined by the swept bandwidth, i.e., $\Delta R \approx c/(2B_s)$, rather than by the pulse duration $T_p$.
As a result, we can choose $T_p$ to be long enough without compromising the range resolution. 
In addition, with fixed $T_p$, the received echo SNR is then determined by the sensing power $P_s$ according to (\ref{se.snr}).

For an LFM sensing pulse with swept bandwidth $B_s$, its squared effective bandwidth is $\beta^2=B_s^2/12$ \cite{skolnik1960}. Substituting this effective bandwidth into \eqref{eq:CRB-tau-general} yields
\begin{equation}
\mathrm{CRB}_{\tau}^{(1)}
=
\frac{3}{2\pi^2}
\frac{N_0}{|\alpha|^2 P_s T_p B_s^2 N}.
\label{eq:CRB-tau-one-pulse}
\end{equation}

Notice that the above derivation is for one sensing pulse. Over the total sensing transmission time $T_s$, multiple $L_s$ coherent pulses provide independent observations of the same quasi-static target parameters after timing and phase compensation. Hence, the accumulated Fisher information is $L_s$ times that of one pulse, and the corresponding CRBs scale inversely with $L_s$ \cite{kay1993}. Accordingly, the delay CRB under {$L_s$} coherent pulses becomes
\begin{equation}
\mathrm{CRB}_{\tau}
=
\frac{3}{2\pi^2}
\frac{N_0}{|\alpha|^2 P_s L_s T_p B_s^2 N}.
\label{eq:CRB-tau}
\end{equation}

Furthermore, via multi-antenna spatial sampling, the MA array can also estimate the AoA $\vartheta$ of the target. The CRB for estimating $\vartheta$ is given by \cite{boyer2010}
\begin{equation}
\mathrm{CRB}_{\vartheta}
=
\frac{N_0\lambda^2}
{16\pi^2 |\alpha|^2 P_s L_s T_p N \cos^2\vartheta\, \mathrm{var}(\bm{x})},
\label{eq:CRB-theta}
\end{equation}
where $\mathrm{var}(\bm{x}) \triangleq\frac{1}{N} \sum_{n=1}^N x_n^2-\left(\frac{1}{N}\sum_{n=1}^N x_n\right)^2$ denotes the spatial variance of the APV.

To obtain the final localization accuracy, the delay CRB and the angle CRB are mapped to the Cartesian coordinate system.
Using the standard CRB transformation rule, the Cartesian position CRB is obtained as
\begin{subequations}
\begin{align}
&\mathrm{CRB}_{\mathrm{pos}}(L_s,B_s,P_s,\bm{x};R,\vartheta)\\
& =
\frac{c^2}{4}\mathrm{CRB}_{\tau}
+
R^2\mathrm{CRB}_{\vartheta}
\label{eq:CRB_two_complement}\\
&=
\frac{N_0}
{8\pi^2 |\alpha|^2 P_s L_s T_p N}
\left[
\frac{3c^2}{B_s^2}
+
\frac{\lambda^2}{2\,\mathrm{var}(\bm{x})}
\cdot
\frac{R^2}{\cos^2\vartheta}
\right]
\label{eq:CRB-pos}\\
&\triangleq
f^{\rm pos}_s(L_s,B_s,P_s,\bm{x};R,\vartheta).
\end{align}
\end{subequations}

The two additive terms in \eqref{eq:CRB-pos} are referred to as the delay-induced and angle-induced components of the position CRB, respectively. The former depends on the sensing bandwidth through $B_s^{-2}$, whereas the latter depends on the APV variance and the target geometry through $\mathrm{var}(\bm{x})^{-1}R^2/\cos^2\vartheta$.

To further account for target uncertainty in a given region $\mathcal C$, the worst-case position CRB over all feasible target locations in $\mathcal C$ is considered.
\begin{proposition}

Define $R_{\max}\triangleq\max_{(R,\vartheta)\in\mathcal C}R$, $\upsilon\triangleq \min_{(R,\vartheta)\in\mathcal C}\cos^2\vartheta>0$, and $\tau_{\max} \triangleq \max_{(R,\vartheta)\in\mathcal C}\frac{2R}{c} = \frac{2R_{\max}}{c}$. For robust sensing, the echo-reception interval is selected to cover the worst-case echo delay, i.e., $T_g\ge \tau_{\max}$, and the number of valid sensing pulses over $\mathcal C$ is
$ L_{s,\mathcal C} \triangleq \frac{T_s}{L T_{\rm slot}}.$ Then, the worst-case position CRB satisfies
\begin{equation}
\max_{(R,\vartheta)\in\mathcal C}
f^{\rm pos}_s(L_s,B_s,P_s,\bm{x};R,\vartheta)
\le
f^{\rm up}_s(T_s,B_s,P_s,\bm{x}),
\end{equation}
where
\begin{equation}
f^{\rm up}_s(T_s,B_s,P_s,\bm{x})
\triangleq
\frac{\chi_1}{L_{s,\mathcal C}P_s B_s^2}
+
\frac{\chi_2}{L_{s,\mathcal C}P_s\,\mathrm{var}(\bm{x})},
\label{eq:CRB-constraint-general}
\end{equation}
with $\chi_1= \frac{3c^2N_0}{8\pi^2|\alpha|^2T_pN}$ and $\chi_2= \frac{N_0\lambda^2R_{\max}^2}{16\pi^2|\alpha|^2T_pN\upsilon}$.
\end{proposition}

\noindent\textit{Proof:} See Appendix A.

To ensure adequate sensing performance, a maximum acceptable CRB threshold $\Gamma_s$ is imposed as
\begin{equation}
f^{\rm up}_s(T_s,B_s,P_s,\bm{x})\le \Gamma_s.
\label{eq:sensing-requirement-general}
\end{equation}

\subsection{Problem Formulation}

The design objective is to maximize the common effective communication rate under a prescribed worst-case sensing accuracy requirement in the TDD \textcolor{black}{or} FDD \textcolor{black}{scheme}.

\begin{figure}[t]
\centering
\includegraphics[width=3.4in,height = 2.0 in]{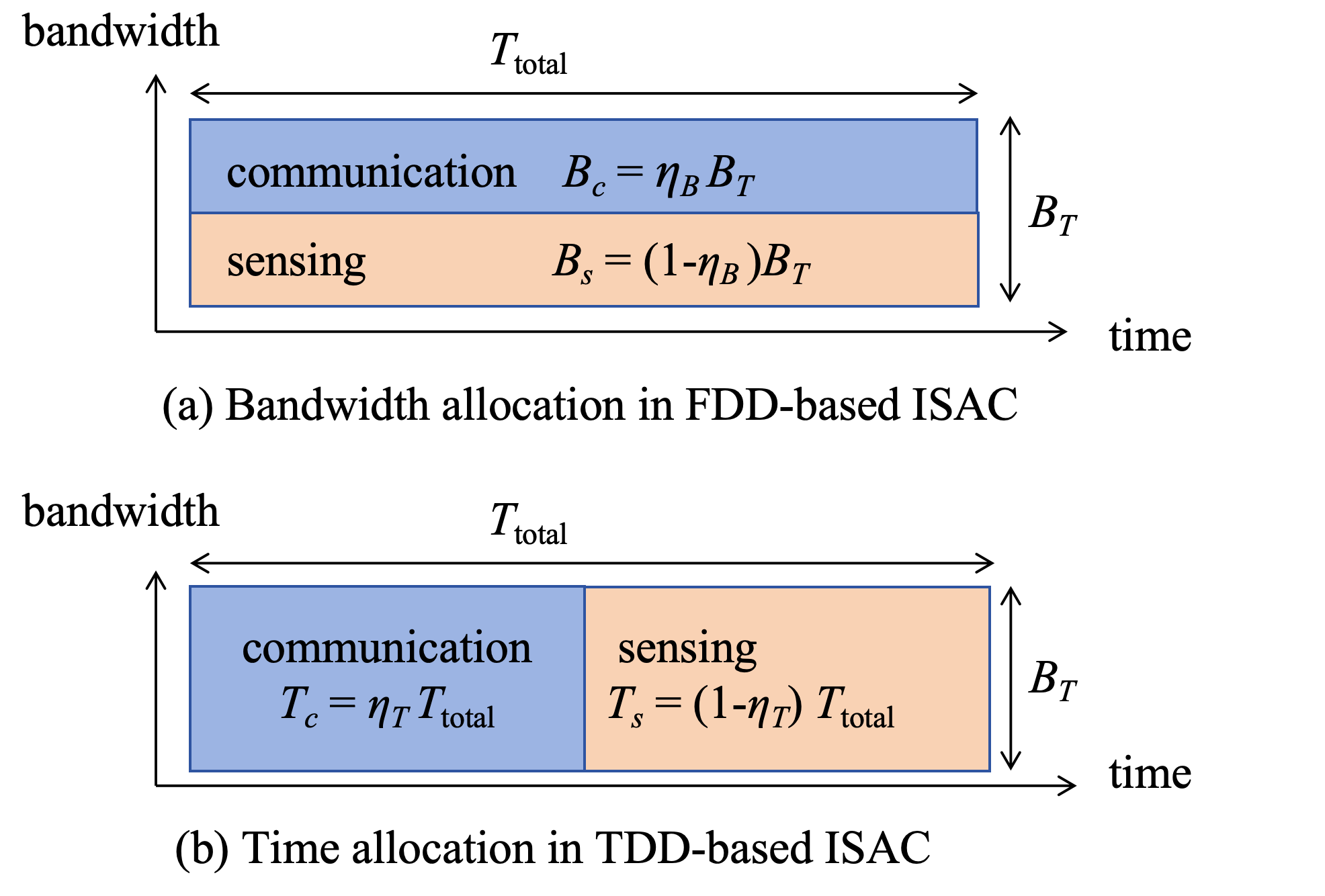}
\caption{\textcolor{black}{Resource allocation schemes in the (a) FDD- and (b) TDD-based ISAC.}}
\label{fig:tdd-fdd-resource-orthogonalization}
\end{figure}

\subsubsection{FDD-Based Formulation}
{As shown in Fig. 3, in the FDD framework, communication and sensing occupy the whole frame over orthogonal frequency bands, so $N_{c,{\rm FDD}}=Q$ and $T_c=T_s=T_{\rm total}$. Let $\eta_B$ and $\eta_P$ denote the bandwidth and power fractions assigned to communication, respectively. Then, the communication bandwidth $B_c$, sensing bandwidth $B_s$, communication power $P_c$ and sensing power $P_s$ are given by}
\begin{align}
&B_c=\eta_B B_T,\quad B_s=(1-\eta_B)B_T, \label{eq:FDD-bandwidth-split}\\
&P_c=\eta_P P_T,\quad P_s=(1-\eta_P)P_T.
\label{eq:FDD-power-split}
\end{align}
The FDD-based common effective rate is
\begin{equation}
f^{\rm ZF}_{c,{\rm FDD}}(\bm x,\eta_B,\eta_P)
=
\eta_B B_T
\log_2
\left(
1+
\tilde{\gamma}(\bm x,\eta_P P_T,\eta_BB_T)
\right).
\label{eq:fc-FDD}
\end{equation}

{Since the whole transmission time is available for sensing in FDD, the number of sensing pulses is $L_{s,{\rm FDD}} \triangleq Q/L$, and the sensing CRB is given as}
\begin{align}
f^{\rm up}_{s,{\rm FDD}}(\eta_B,\eta_P,\bm x)
&=
\frac{\chi_1}
{L_{s,{\rm FDD}}(1-\eta_P)P_T[(1-\eta_B)B_T]^2}
\nonumber\\
&\quad+
\frac{\chi_2}
{L_{s,{\rm FDD}}(1-\eta_P)P_T\,\mathrm{var}(\bm x)}.
\label{eq:fs-FDD}
\end{align}
{The FDD-based communication-sensing tradeoff problem is formulated as}
\begin{subequations}
\begin{align}
(\rm{P1\mbox{-}FDD}):\quad
& \max_{\eta_B,\eta_P,\bm{x}}
\, f^{\rm ZF}_{c,{\rm FDD}}(\bm x,\eta_B,\eta_P)
\\
\text{s.t.}\quad
& f^{\rm up}_{s,{\rm FDD}}(\eta_B,\eta_P,\bm x) \le \Gamma_s,
\label{eq:FDD-constraint-sensing}\\
& x_1 \ge 0,\; x_N \le D,
\label{eq:FDD-constraint-region}\\
& x_n-x_{n-1}\ge d_{\min},\quad n=2,3,\ldots,N,
\label{eq:FDD-constraint-distance}\\
& 0\le \eta_B\le 1,
\label{eq:FDD-constraint-etaB}\\
& 0\le \eta_P\le 1.
\label{eq:FDD-constraint-etaP}
\end{align}
\end{subequations}

\subsubsection{TDD-Based Formulation}

As shown in Fig. 3, in TDD, let $N_{s,{\rm TDD}}\in\mathbb Z^{+}$ denote the number of sensing pulses in the total sensing transmission time. Since each sensing pulse cycle occupies $L$ time slots, $LN_{s,{\rm TDD}}$ slots are allocated to sensing, and the number of remaining slots allocated to communication is $N_{c,{\rm TDD}} =Q-LN_{s,{\rm TDD}}$.
The corresponding communication time fraction is defined as
\begin{equation}
\eta_T
\triangleq
\frac{N_{c,{\rm TDD}}}{Q}
=
\frac{Q-LN_{s,{\rm TDD}}}{Q}.
\label{eq:TDD-time-fraction}
\end{equation}
During their active intervals, both functions use the full system bandwidth and transmit power, i.e.,
\begin{align}
& B_c=B_s=B_T,\quad \quad P_c=P_s=P_T. \label{eq:TDD-power-split}
\end{align}
Based on the above derivation, the TDD-based common effective rate is
\begin{equation}
f^{\rm ZF}_{c,{\rm TDD}}(\bm x,N_{s,{\rm TDD}})
=
\eta_T B_T
\log_2
\left(
1+
\tilde{\gamma}(\bm x,P_T,B_T)
\right).
\label{eq:fc-TDD}
\end{equation}
The corresponding sensing CRB upper bound is
\begin{align}
f^{\rm up}_{s,{\rm TDD}}(N_{s,{\rm TDD}},\bm x)=\frac{1}{N_{s,{\rm TDD}}P_T}\big[\frac{\chi_1}
{B_T^2}+\frac{\chi_2}{\mathrm{var}(\bm x)}\big].
\label{eq:fs-TDD}
\end{align}
Therefore, the TDD-based communication-sensing tradeoff problem is formulated as
\begin{subequations}
\begin{align}
(\rm{P1\mbox{-}TDD}):\quad
& \max_{N_{s,{\rm TDD}},\bm{x}}
\, f^{\rm ZF}_{c,{\rm TDD}}(\bm x,N_{s,{\rm TDD}})
\\
\text{s.t.}\quad
& f^{\rm up}_{s,{\rm TDD}}(N_{s,{\rm TDD}},\bm x) \le \Gamma_s,
\label{eq:TDD-constraint-sensing}\\
& x_1 \ge 0,\; x_N \le D,
\label{eq:TDD-constraint-region}\\
& x_n-x_{n-1}\ge d_{\min},\quad n=2,3,\ldots,N.
\label{eq:TDD-constraint-distance}\\
& N_{s,{\rm TDD}} \in\mathbb Z^{+},
\label{eq:TDD-constraint-pulse}\\
& LN_{s,{\rm TDD}}\le Q.
\label{eq:TDD-constraint-slot}
\end{align}
\end{subequations}
{In (P1-FDD) and (P1-TDD), \eqref{eq:FDD-constraint-sensing} and \eqref{eq:TDD-constraint-sensing}, respectively, enforce the prescribed \textcolor{black}{sensing} CRB requirement. The constraints \eqref{eq:FDD-constraint-region}, \eqref{eq:FDD-constraint-distance}, \eqref{eq:TDD-constraint-region}, and \eqref{eq:TDD-constraint-distance} impose the common APV movement region and minimum spacing requirements. Finally, \eqref{eq:FDD-constraint-etaB} and \eqref{eq:FDD-constraint-etaP} specify the bandwidth and power allocation in FDD, whereas \eqref{eq:TDD-constraint-pulse} and \eqref{eq:TDD-constraint-slot} govern the sensing pulse number and time allocation in TDD.}

\section{Proposed Algorithm for FDD-Based Formulation}
Here, we first derive the optimization algorithm for the FDD-based problem. To address the coupling among the bandwidth allocation coefficient $\eta_B$, the power allocation coefficient $\eta_P$, and the APV $\bm{x}$, we adopt an alternating optimization (AO) algorithm to obtain a locally optimal solution for problem (P1-{\rm FDD}). Specifically, the proposed method updates $(\eta_B,\eta_P)$ and $\bm{x}$ in an iterative manner, as detailed below.

\subsection{Optimization of $\eta_B$ and $\eta_P$ with Fixed $\bm{x}$}
In the AO procedure, when the APV $\bm{x}$ is fixed, the joint optimization over the bandwidth allocation coefficient $\eta_B$ and the power fraction $\eta_P$ reduces to the following problem\textcolor{black}{,}
\begin{subequations}
\label{eq:P_eta}
\begin{align}
(\mathrm{P2\mbox{-}FDD}): \max_{\eta_B,\eta_P} \quad &\eta_B \log_2\!\left(1+\kappa\,\frac{\eta_P}{\eta_B}\right)
\label{eq:P_eta_obj}\\
\text{s.t.}\quad & \frac{c_1/(1-\eta_B)^2+c_2}{1-\eta_P}-\Gamma_s \le 0,
\label{eq:P_eta_crb}\\
& 0\le \eta_B \le 1,
\label{eq:P_eta_boxB}\\
& 0\le \eta_P \le 1,
\label{eq:P_eta_boxP}
\end{align}
\end{subequations}
where the notations are defined as
\begin{align}
&\kappa \triangleq \frac{P_T}{\mathrm{tr}\!\big[\big(\bm H(\bm{x},\bm\theta)^H \bm H(\bm{x},\bm\theta)\big)^{-1}\big]\;N_0\;B_T}, \nonumber \\
& c_1 \triangleq \frac{\chi_1}{L_{s,{\rm FDD}} P_T B_T^2}, \qquad c_2 \triangleq \frac{\chi_2}{L_{s,{\rm FDD}} P_T\,\mathrm{var}(\bm{x})}. \label{eq:kakka}
\end{align}
It is worth noting that, for a fixed APV $\bm{x}$, the subproblem with respect to $(\eta_B,\eta_P)$ is a convex optimization problem. Specifically, the objective function $\eta_B\log_2 \left(1+\kappa\,\eta_P/\eta_B\right)$ is jointly concave in $(\eta_B,\eta_P)$, since it is the perspective of a concave $\log(\cdot)$ function, and the constraint \eqref{eq:P_eta_crb} is jointly convex over $\eta_B$ and $\eta_P$. Therefore, the optimal solution can be obtained from the KKT conditions.
\subsection{Optimization of $\bm{x}$ with Fixed $\eta_B$ and $\eta_P$}
With fixed bandwidth and power allocation variables, the APV subproblem is written as
\begin{subequations}
\begin{align}
(\mathrm{P3\mbox{-}FDD}): \quad \min_{\bm{x}}\quad & \mathcal J(\bm{x}) \triangleq \mathrm{tr}\big[\big(\bm H(\bm{x})^{H}\bm{H}(\bm{x})\big)^{-1}\big] \\
\text{s.t.}\quad & g(\bm{x}) \triangleq \bm{x}^{T}\bm{B}\bm{x}\geq \chi_4,
\label{eq:var-x}\\
& \text{(\ref{eq:FDD-constraint-region})}, \text{(\ref{eq:FDD-constraint-distance})},
\end{align}
\end{subequations}
where we define
\begin{align}
& \bm{B} \triangleq \frac{1}{N} \bm{I}_{N} - \frac{1}{N^2}\bm{1}_{N}\bm{1}^T_{N}, \\
& \chi_{4} \triangleq
\frac{\chi_2}
{L_{s,{\rm FDD}}\Gamma_s(1-\eta_P)P_T
-\frac{\chi_1}{(1-\eta_B)^2B_T^2}}.
\end{align}
The objective function $\mathcal J(\bm{x})$ is non-convex with respect to the variable $\bm{x}$. To address this problem, we employ a projected-gradient method. To obtain the descent direction, we first derive the gradient of $\mathcal J(\bm{x})$ as follows.
\begin{proposition}
\label{prop:gradient} Define $\mathcal J(\bm{x}) \triangleq \mathrm{tr}\!\left(\big(\bm{H}(\bm{x})^{H}\bm{H}(\bm{x})\big)^{-1}\right)$, $k_\lambda\triangleq 2\pi/\lambda$, $\bm S(\bm x)\triangleq \bm H(\bm x)^H\bm H(\bm x)$, $\bm G(\bm x)\triangleq \bm H(\bm x)\bm S(\bm x)^{-2}$, and $\bm c\triangleq[\sin\theta_1,\ldots,\sin\theta_K]^T$. Then, $\mathcal J(\bm{x})$ is differentiable with respect to $\bm{x}$, and its gradient is given by
\begin{equation}
\nabla \mathcal J(\bm{x}) = 2k_\lambda\,{\rm{Im}} \left((\overline{\bm{G}(\bm{x})}\odot \bm{H}(\bm{x}))\,\bm{c}\right).
\end{equation}
\end{proposition}
\noindent\textit{Proof:} See Appendix B.

To handle the non-convex constraint (\ref{eq:var-x}), we apply successive convex approximation (SCA) by replacing $g(\bm{x})$ at the $p$-th iteration with its first-order lower bound at $\bm{x}^{(p)}$, which is obtained as
\begin{align}
g(\bm{x}) & \geq   2 (\bm{x}^{(p)})^T \bm{B} \bm{x} -g(\bm{x}^{(p)}). 
\end{align}
{This transformation yields the half-space constraint
\begin{equation}
\mathcal{C}_1^{(p)} \triangleq
\left\{\bm{x}\in\mathbb{R}^{N}:\ \bm{u}^{(p)T}\bm{x}\ge b^{(p)}\right\},
\end{equation}
where $\bm{u}^{(p)}\triangleq 2\bm{B}\bm{x}^{(p)}$ and $b^{(p)}\triangleq \chi_4+g(\bm{x}^{(p)})$. Together with the physical movement and minimum-spacing set
\begin{align}
\mathcal{C}_2 \triangleq \{\bm{x}\in\mathbb{R}^{N}:&\ x_1\ge 0,\ x_N\le D, \nonumber \\
&\ x_n-x_{n-1}\ge d_{\min}, n=2,\ldots,N \},
\label{eq:C1C2_def}
\end{align}
the projection in the $p$-th iteration is performed over $\mathcal{C}^{(p)}\triangleq \mathcal{C}_1^{(p)}\cap\mathcal{C}_2$.}

Given the current iterate $\bm{x}^{(p)}$, one projected-gradient update with step size $\varepsilon>0$ is
\begin{equation}
\bm{x}^{(p+1)}(\varepsilon) = \Pi_{\mathcal{C}^{(p)}}\!\left( \bm{x}^{(p)}-\varepsilon \nabla \mathcal J(\bm{x}^{(p)}) \right),
\label{eq:pg_update_core_rewrite}
\end{equation}
where $\Pi_{\mathcal{A}}(\cdot)$ denotes the Euclidean projection onto a closed convex set $\mathcal{A}$. The step size $\varepsilon$ is selected by backtracking line search. Starting from $\varepsilon_{\mathrm{init}}>0$, we reduce $\varepsilon$ as $\varepsilon\leftarrow\bar{\rho}\varepsilon$ with $\bar{\rho}\in(0,1)$ until the Armijo condition
\begin{equation}
\mathcal J\!\left(\bm{x}^{(p+1)}(\varepsilon)\right) \le \mathcal J(\bm{x}^{(p)})-\bar{\kappa} \varepsilon\|\nabla \mathcal J(\bm{x}^{(p)})\|_2^{2},
\label{eq:armijo_cond}
\end{equation}
is satisfied, where $\bar{\kappa}\in(0,1)$ is a constant. The projected-gradient iterations stop when the relative decrease of $\mathcal J(\bm{x})$ is below $\epsilon_{\rm pg}$ or after $T_{\rm in}$ iterations.

The main computational task in \eqref{eq:pg_update_core_rewrite} is the projection onto $\mathcal{C}^{(p)}$. Instead of solving this projection by a generic convex solver, we exploit the structure of $\mathcal{C}^{(p)}=\mathcal{C}_1^{(p)}\cap\mathcal{C}_2$. The set $\mathcal{C}_1^{(p)}$ is a half-space and has a closed-form projection. The projection onto $\mathcal{C}_2$ can be converted into a bounded isotonic regression problem after the variable shift $y_n=x_n-(n-1)d_{\min}$, and is efficiently solved by the pool-adjacent-violators (PAV) algorithm \cite{Barlow_1972}. With these two basic projections, Dykstra's algorithm \cite{Dykstra_1983} is used to compute the Euclidean projection onto their intersection $\mathcal{C}^{(p)}$. Finally, the Armijo condition in \eqref{eq:armijo_cond} guarantees that $\mathcal J(\bm{x})$ is non-increasing over the projected-gradient iterations, while $\mathcal J(\bm{x})$ is lower-bounded by zero.

As a summary, the complete algorithmic flow is given in Algorithm~\ref{alg:overall_AO}. In each outer AO iteration indexed by $r$, the APV is refined by an inner projected-gradient loop indexed by $\ell$, where $\bm{x}^{(r,\ell)}$ denotes the APV after the $\ell$-th inner update and $\mathcal C^{(r,\ell)}$ denotes the convex feasible set constructed at $\bm{x}^{(r,\ell)}$. The alternating algorithm is repeated until the relative improvement of the objective value is below $\epsilon_{\rm th}$ or the maximum number of outer iterations $T_{\rm out}$ is reached. For the overall alternating procedure, the objective value of (P1-FDD) is non-decreasing after the resource allocation and APV update steps and is upper-bounded by a finite value under finite transmit power, finite bandwidth, and physically nonzero propagation distances. Hence, Algorithm~\ref{alg:overall_AO} yields a convergent sequence of objective values.
\begin{algorithm}[!t]
\caption{Alternating Optimization for Solving Problem (P1-{\rm FDD})}
\label{alg:overall_AO}
\begin{algorithmic}[1]
\STATE \textbf{Input:} $\bm{x}^{(0)}$, $T_{\rm out}$, $T_{\rm in}$, $\epsilon_{\rm th}$, $\epsilon_{\rm pg}$, $\varepsilon_{\rm init}$, $\bar{\rho}$, and $\bar{\kappa}$
\STATE Set $r=0$
\WHILE{$r<T_{\rm out}$}
\STATE Obtain $(\eta_B^{(r+1)},\eta_P^{(r+1)})$ by solving problem (P2-FDD) with fixed $\bm{x}^{(r)}$
\STATE Set $\ell=0$ and $\bm{x}^{(r,0)}=\bm{x}^{(r)}$
\WHILE{$\ell<T_{\rm in}$}
\STATE Set $\varepsilon=\varepsilon_{\rm init}$
\STATE $\tilde{\bm{x}}^{(r,\ell+1)} = \Pi_{\mathcal{C}^{(r,\ell)}}\!\left( \bm{x}^{(r,\ell)} - \varepsilon\nabla \mathcal J(\bm{x}^{(r,\ell)}) \right)$ by Dykstra projection
\WHILE{$\mathcal J(\tilde{\bm{x}}^{(r,\ell+1)}) > \mathcal J(\bm{x}^{(r,\ell)}) - \bar{\kappa} \varepsilon \left\|\nabla \mathcal J(\bm{x}^{(r,\ell)})\right\|_2^2$}
\STATE $\varepsilon\leftarrow\bar{\rho}\varepsilon$
\STATE $\tilde{\bm{x}}^{(r,\ell+1)} = \Pi_{\mathcal{C}^{(r,\ell)}}\!\left( \bm{x}^{(r,\ell)} - \varepsilon\nabla \mathcal J(\bm{x}^{(r,\ell)}) \right)$ by Dykstra projection
\ENDWHILE
\STATE $\bm{x}^{(r,\ell+1)}=\tilde{\bm{x}}^{(r,\ell+1)}$
\IF{$|\mathcal J(\bm{x}^{(r,\ell+1)})-\mathcal J(\bm{x}^{(r,\ell)})|\le\epsilon_{\rm pg}$}
\STATE $\ell\leftarrow\ell+1$
\STATE \textbf{break}
\ENDIF
\STATE $\ell\leftarrow\ell+1$
\ENDWHILE
\STATE Set $\bm{x}^{(r+1)}=\bm{x}^{(r,\ell)}$
\STATE $r\leftarrow r+1$
\IF{$|f^{\rm ZF}_{c, \rm {FDD}}(\bm{x}^{(r)},\eta_B^{(r)},\eta_P^{(r)})-f^{\rm ZF}_{c, {\rm FDD}}(\bm{x}^{(r-1)},\eta_B^{(r-1)},\eta_P^{(r-1)})|\le\epsilon_{\rm th}$}
\STATE \textbf{break}
\ENDIF
\ENDWHILE
\STATE \textbf{Output:} $(\eta_B,\eta_P,\bm{x}^\star)$
\end{algorithmic}
\end{algorithm}

\vspace{-0.2cm}
\subsection{Complexity Analysis}
We next analyze the computational complexity of the proposed APV optimization. In each projected-gradient iteration, the gradient evaluation and the objective evaluations required by the line search each cost $\mathcal O(NK^2+K^3)$. The proposed low-complexity projected-gradient method with Dykstra-based projection has an overall complexity of $\mathcal O \left(T_{\rm in}\left[(1+T_{\rm ls})(NK^2+K^3)+T_{\rm ls}T_{\rm dyk}N\right]\right)$, where $T_{\rm ls}$, $T_{\rm in}$, and $T_{\rm dyk}$ denote the numbers of line-search trials, projected-gradient iterations, and Dykstra iterations, respectively. For comparison, under the same projected-gradient procedure, each convex projection subproblem can instead be solved by CVX, yielding an overall complexity of $\mathcal O \left(T_{\rm in}\left[(1+T_{\rm ls})(NK^2+K^3)+T_{\rm ls}T_{\rm cvx}N^3\right]\right)$, where $T_{\rm cvx}$ denotes the number of interior-point iterations. Therefore, the projection-related cost of the proposed method is linear in $N$, compared with the cubic-order CVX baseline.

\section{Proposed Algorithm for the TDD-Based Formulation}

In the TDD-based optimization problem (P1-{\rm TDD}), the sensing pulse number $N_{s,{\rm TDD}}$ and the APV $\bm{x}$ are intrinsically coupled. A smaller $N_{s,{\rm TDD}}$ preserves more time slots for communication, but it also tightens the sensing requirement on $\mathrm{var}(\bm{x})$, thereby reducing the freedom of APV optimization for improving the multiuser communication channels. Conversely, a larger $N_{s,{\rm TDD}}$ sacrifices communication time but relaxes the APV variance requirement.

{Since $N_{s,{\rm TDD}}$ is a bounded integer variable satisfying $LN_{s,{\rm TDD}}\le Q$, we adopt a nested procedure: the outer layer searches over the sensing pulse number $N_{s,{\rm TDD}}$, while the inner layer optimizes the APV $\bm{x}$ for each candidate by using the method developed in Subsection~III-B.} We first determine the feasible search interval. Let $V_{\max}$ denote the maximum APV variance achievable under the antenna movement and minimum-spacing constraints \cite{MA_sensing}. 
Substituting this sensing-favorable APV variance into \eqref{eq:fs-TDD} and neglecting integer rounding of the sensing pulse number, the lower bound of the sensing pulse number $N_{s,{\rm TDD}}$ is given as
\begin{equation}
N_{s,{\rm lb}}
=\frac{\chi_1}{B_T^2 P_T \Gamma_s}+\frac{\chi_2}{V_{\max}P_T\Gamma_s}.
\label{eq:Ns-lb}
\end{equation}
Meanwhile, the number of sensing pulses cannot exceed the total frame duration. Therefore, the upper bound follows from \eqref{eq:TDD-constraint-slot} as $N_{s,{\rm ub}} = \frac{Q}{L}$.
The candidate sensing pulse numbers form the set $\mathcal{N}_{s,{\rm TDD}} = \{N_{s,{\rm lb}},N_{s,{\rm lb}}+1,\ldots,N_{s,{\rm ub}}\}$.
For any fixed $N_{s,{\rm TDD}}\in\mathcal{N}_{s,{\rm TDD}}$, the APV subproblem is thus transformed as
\begin{subequations}
\label{eq:TDD-APV-subproblem}
\begin{align}
\!\!\!\!(\rm{P4\mbox{-}TDD}):  \max_{\bm{x}}\,\,\,
& \eta_T B_T \log_2 \left( 1+ \tilde{\gamma}(\bm{x},P_T,B_T) \right) \label{eq:TDD-APV-subproblem-obj}\\[2pt]
\mathrm{s.t.}\quad
& \mathrm{var}(\bm{x})\ge V_{\rm req}(N_{s,{\rm TDD}}),
\label{eq:TDD-APV-subproblem-var}\\[2pt]
& x_1 \ge 0,\quad x_N\le D,
\label{eq:TDD-APV-subproblem-region}\\[2pt]
& x_n-x_{n-1}\ge d_{\min},\quad n=2,\ldots,N,
\label{eq:TDD-APV-subproblem-spacing}
\end{align}
\end{subequations}
where $V_{\rm req}(N_{s,{\rm TDD}}) \triangleq \frac{B_T^2 \chi_2} {B_T^2 N_{s,{\rm TDD}} P_T\Gamma_s - \chi_1}$. This problem can be solved using the projected-gradient method with SCA and Dykstra projection developed in Subsection~III-B. After solving the APV subproblem for all candidate sensing pulse numbers, the solution achieving the largest communication rate is selected.

The inner APV optimization (P4-{\rm TDD}) has the same complexity as (P3-{\rm FDD}) as described in Subsection~III-C. Since the TDD scheme solves this APV subproblem for each feasible sensing pulse number, its additional computational cost is proportional to the number of candidate pulse numbers.

\section{Performance Analysis of MA-Aided TDD/FDD ISAC}

\label{sec:concise-architecture-analysis-added}

This section characterizes the TDD/FDD performance under different communication and sensing conditions and investigates how the role of MA varies across representative operating regimes through its effects on multiuser channel orthogonality and sensing aperture.


\label{subsec:ma-scenarios-added}

The APV affects communication and sensing through different mechanisms. 
On the communication side, the user angular distribution determines how strongly the multiuser channel orthogonality depends on the antenna positions. When the users are angularly well separated, i.e., their angles $\{\theta_k\}_{k=1}^{K}$ have relatively large pairwise separations, favorable channel orthogonality can generally be maintained over a relatively broad range of APV configurations. In contrast, when the users are angularly clustered, i.e., their angles $\{\theta_k\}_{k=1}^{K}$ are concentrated within a relatively narrow angular region, their steering vectors are more strongly correlated, and the APV configuration becomes more important for improving channel orthogonality.
To quantify the impact of the APV optimization on the communication performance, we define the normalized SNR $q(\bm{x})=\kappa(\bm{x})/\kappa_\perp$, where
\begin{equation}
\kappa(\bm{x}) = \frac{P_T}{N_0 B_T\,\mathrm{tr}[(\bm H^H\bm H)^{-1}]},
\,
\kappa_\perp = \frac{N P_T}{N_0 B_T \sum_{k=1}^{K}|\beta_k|^{-2}}.
\end{equation}
Here, $\kappa(\bm{x})$ is the communication SNR evaluated at $P_c=P_T$ and $B_c=B_T$, while $\kappa_\perp$ is the SNR upper bound attained with mutually orthogonal user channels. A value of $q$ near one indicates little SNR loss due to channel correlation, whereas a value well below one indicates a substantial loss. Therefore, $q$ is used to characterize how effectively MA improves the communication-channel orthogonality under different user angle configurations.

For sensing, define the delay- and angle-related loads under full sensing bandwidth, power and time slots, i.e., $B_s = B_T,P_s = P_T$ and $T_s = T_{\rm total}$ as
\begin{equation}
d=\frac{\chi_1}{L_{s,{\rm FDD}}P_TB_T^2\Gamma_s},
\,
a(\bm{x})=\frac{\chi_2}{L_{s,{\rm FDD}}P_T\Gamma_s\,\mathrm{var}(\bm{x})},
\end{equation}
where $L_{s,{\rm FDD}} = Q/L$ is the number of sensing pulses available if the whole frame is used for sensing. We assume $0\le d<1$ and $0\le a<1$, respectively, which corresponds to a feasible sensing requirement with a nonzero amount of resources remaining for communication. In addition, their ratio is defined as
$\xi(\bm{x})=\frac{a(\bm{x})}{d} =\frac{\chi_2B_T^2}{\chi_1\mathrm{var}(\bm{x})}$.
To facilitate the analysis, let $\xi\ll1$ and $\xi\gg1$ correspond to delay-dominant and angle-dominant sensing under the full system bandwidth, respectively. 
Under the antenna movement and minimum-spacing constraints, the APV variance satisfies $V_{\min}\le\mathrm{var}(\bm{x})\le V_{\max}$, where
\begin{align}
V_{\min} & = \frac{N^2-1}{12}d_{\min}^2, \\
V_{\max} & = V_{\min}+\frac{\lfloor N^2/4\rfloor}{N^2}\big[D-(N-1)d_{\min}\big](D+d_{\min}). \nonumber
\end{align}
The lower bound $V_{\min}$ is attained by packing adjacent antennas at the minimum spacing, while the upper bound $V_{\max}$ is attained by splitting the antennas as evenly as possible into two groups packed toward the two boundaries. Consequently, APV adjustment can change the relative contributions of the angle-/delay-related CRB components within this feasible range. 

Based on the user angular distribution and the sensing-load composition, the role of MA can be interpreted through four representative scenarios.

\textbf{Scenario 1: Angularly well-separated users and delay-dominant sensing.}
For angularly well-separated users, favorable channel orthogonality can be achieved over a relatively broad range of APVs, so there is limited need to adjust the antenna positions specifically for communication-channel decorrelation. Under delay-dominant sensing, enlarging the antenna position variance also provides limited sensing benefit because the dominant delay-related term is unaffected by the sensing aperture. 


The principal difference between the two schemes lies in the sensing-bandwidth cost. The delay-related CRB term scales inversely with the square of the sensing bandwidth.
This comparison can be made explicit by considering the extreme case with the pure delay-sensing limit, i.e., $a=0$.
Neglecting integer rounding of the sensing pulse number, 
the achievable communication rates in the two schemes are, respectively, given as
\begin{align}
&R_{\rm TDD}=B_T(1-d)\log_2(1+\kappa), \\
&R_{\rm FDD}=B_T\eta_B\log_2(1+\kappa\frac{\eta_P}{\eta_B}).
\end{align}
Let $\delta_B$ and $\delta_P$ denote the sensing bandwidth and sensing power fractions in FDD, respectively. The sensing constraint requires $\delta_P\delta_B^2\geq d$, which gives $\delta_B\geq\sqrt{d}$ and $\delta_P\geq d$. Accordingly, the communication bandwidth and power fractions in FDD satisfy $\eta_B\leq1-\sqrt{d}\leq1-d$ and $\eta_P\leq1-d$. 
Notice that the communication rate is monotonically increasing with respect to both $\eta_B$ and $\eta_P$. Since both communication bandwidth/power fractions in FDD are no larger than the communication time fraction $\eta_T \triangleq 1-d$ in TDD, it follows that
$R_{\rm FDD}\leq B_T(1-d)\log_2(1+\kappa)=R_{\rm TDD}$.
Although the above rate comparison assumes a common APV, the ordering $R_{\rm FDD}\leq R_{\rm TDD}$ holds for every feasible APV in the pure delay-sensing case. It therefore remains valid after independent APV optimization, even if the two schemes select different optimal APVs.

For the general delay-dominant regime with $0<a\ll d$, the optimized APVs of the two schemes may differ. Nevertheless, for angularly well-separated users, $\kappa(\bm{x})$ is relatively insensitive to the APV over a broad range of configurations. Hence, the communication SNR difference induced by the two optimized APVs is expected to be modest, and TDD tends to retain an advantage owing to its full sensing bandwidth, although a universal rate ordering is no longer guaranteed.

\textbf{Scenario 2: Angularly well-separated users and angle-dominant sensing.}
As in Scenario~1, APV optimization is expected to have only a modest effect on $\kappa$ for angularly well-separated users, while the angle-dominant sensing condition favors a larger antenna position variance.

To clarify the comparison between the two schemes, we still first consider the extreme case with pure angle-sensing limit, i.e., $d=0$. In this case, the sensing performance is independent of the sensing bandwidth, so FDD can allocate nearly the full system bandwidth to communication. Under the same sensing requirement, the total communication energy over one frame is the same for the two schemes, i.e.,
$E_{{\rm c,TDD}}=E_{{\rm c,FDD}}=(1-a)P_TT_{\rm total}$.
The difference is that TDD concentrates this energy over a shorter communication interval, while FDD spreads it over the entire frame. By the concavity of the Shannon rate with respect to the communication power $P_c$,
FDD achieves a rate no lower than TDD in the pure angle-sensing limit.
The above comparison is made for a common APV. Since the rate ordering holds for every feasible APV when $d=0$, the same ordering remains valid even when TDD and FDD independently optimize their APVs.

For the general angle-dominant sensing regime with $0< d\ll a$, a sensing condition that is angle dominant under the full system bandwidth does not necessarily remain angle dominant after FDD resource allocation. The effective angle-to-delay CRB ratio after FDD resource partitioning is
\begin{equation}
\widetilde{\xi}_{\rm FDD}
=\frac{a/\delta_P}{d/(\delta_P\delta_B^2)}
=\xi\delta_B^2.
\end{equation}
Therefore, even if the sensing condition under full bandwidth satisfies $\xi>1$, FDD bandwidth partitioning can reduce the effective ratio below one, i.e., $\widetilde{\xi}_{\rm FDD}<1$, so that the resulting sensing-load composition becomes effectively delay dominated. The sensing-power fraction does not affect this ratio because it scales the angle- and delay-related components equally. Similarly, sensing-time allocation in TDD scales both components equally, and hence $\widetilde{\xi}_{\rm TDD}=\xi$.
In addition, APV optimization can reduce \(a(\bm{x})\) by enlarging the antenna position variance, while FDD bandwidth partitioning further reduces \(\widetilde{\xi}_{\rm FDD}\) through the factor \(\delta_B^2\). Consequently, the effective sensing-load composition may shift toward delay dominance, weakening the advantage of FDD and making the full-bandwidth sensing benefit of TDD increasingly relevant.

\textbf{Scenario 3: Angularly clustered users and delay-dominant sensing.}
For angularly clustered users, the communication channels are more strongly correlated and become more sensitive to the APV configuration. MA can therefore improve multiuser channel orthogonality through antenna position adjustment. 

Under delay-dominant sensing, TDD benefits from using the full system bandwidth for sensing and is generally more favorable than FDD.

\textbf{Scenario 4: Angularly clustered users and angle-dominant sensing.}
For angularly clustered users, the communication channel orthogonality is sensitive to the APV configuration, while angle-dominant sensing requirement favors a larger antenna position variance. Therefore, APV optimization needs to balance communication channel orthogonality and sensing aperture.  

If the APV DoF is mainly used to improve communication channel orthogonality and the sensing condition remains strongly angle dominated, FDD can retain its rate advantage. In contrast, if APV adjustment substantially enlarges the sensing aperture, together with FDD bandwidth partitioning, this can further reduce \(\widetilde{\xi}_{\rm FDD}\) and shift the effective sensing-load composition toward delay dominance, making the full-bandwidth sensing benefit of TDD more relevant. Therefore, neither scheme has a universal rate advantage in this regime.

\vspace{-0.2cm}
\section{Numerical Results}
\subsection{Simulation Setup}
This subsection specifies the simulation setup. Unless otherwise specified, the carrier frequency is set to $f_0=26$ GHz. The BS employs $N=8$ movable receive antennas to serve $K=4$ single-antenna users. The antennas move over a one-dimensional region of length $D=0.5$ m with minimum adjacent spacing $d_{\min}=\lambda/4$. The total bandwidth and transmit power are $B_T=15$ MHz and $P_T=6$ W, respectively, and the noise power spectral density is $N_0=10^{-18}$ W/Hz. The user angles are uniformly spaced over $[-\pi/2,\pi/2]$, and their distances are generated within $[30,60]$ m.

For sensing, the pulse width is $T_p=5~\mu\text{s}$ and the target range is $R_{\max}=2250$ m. The echo-reception interval is set to $T_g=2R_{\max}/c=15~\mu\text{s}$. We set the frame duration and time-slot duration to $T_{\rm total}=2$ ms and $T_{\rm slot}=10~\mu\text{s}$, respectively, yielding $Q=200$ slots per frame. Each sensing pulse occupies $L=2$ slots, and hence $L_{s,{\rm FDD}}=100$ sensing pulses are available to FDD in each frame. The target reflection coefficient is modeled by the constant magnitude $|\alpha|=10^{-7}$. Unless otherwise specified, the sensing threshold is $\Gamma_s=4\times10^{-5}$. The APV optimization uses at most $T_{\rm out}=5$ outer iterations and $T_{\rm in}=100$ inner iterations.

To demonstrate the effectiveness of the proposed design, the following benchmark schemes are considered for comparison:
\begin{itemize}
\item \textbf{Proposed joint optimization}: The APV and the scheme-specific resource variables are jointly optimized to maximize the minimum achievable communication rate.

\item \textbf{Sparse uniform linear array (ULA-sparse)}: The antennas are fixed with uniform spacing over $[0,D]$, while only the scheme-specific resource variables are optimized.

\item \textbf{Dense ULA (ULA-dense)}: The antennas are fixed with spacing $d_{\rm min}$, while only the scheme-specific resource variables are optimized.

\item \textbf{Communication-only optimization}: All resources are assigned to communication, and the APV is optimized for the maximum minimum achievable rate.

\item \textbf{Sensing-only optimization}: All resources are assigned to sensing, using the sensing-optimal APV in \cite{MA_sensing}.

\item \textbf{Sensing-optimal APV}: The APV is fixed to the sensing-optimal design in \cite{MA_sensing}, and only the scheme-specific resource variables are optimized for communication.

\item \textbf{Communication-optimal APV}: The APV is first numerically optimized to maximize $\kappa(\bm{x})$ without a sensing constraint. The resulting APV is then fixed and only the scheme-specific resource variables are optimized for communication.
\end{itemize}

Here, scheme-specific resource allocation refers to the sensing/communication time fractions under the TDD scheme and the sensing/communication bandwidth and power fractions under the FDD scheme.

\vspace{-0.1cm}
\subsection{FDD-Based Performance}
We first evaluate the proposed FDD-based MA-enhanced ISAC design.

\begin{table}[t]
\caption{Runtime \textcolor{black}{(In Second)} comparison of the APV optimization for the proposed method and the CVX-based projection baseline.}
\label{tab:apv_runtime}
\centering
\scriptsize
\setlength{\tabcolsep}{3pt}
\begin{tabular*}{\columnwidth}{@{\extracolsep{\fill}}lccccc@{}}
\toprule
Method & \makecell{$K=2$\\$N=8$} & \makecell{$K=4$\\$N=8$} & \makecell{$K=2$\\$N=16$} & \makecell{$K=4$\\$N=16$} & \makecell{$K=8$\\$N=16$} \\
\midrule
CVX & 7.0941 & 8.3968 & 8.0555 & 9.0811 & 10.115 \\
Proposed & 0.040279 & 0.041812 & 0.041936 & 0.044849 & 0.054201 \\
\bottomrule
\end{tabular*}
\end{table}

Table~\ref{tab:apv_runtime} reports the mean MATLAB runtime of one complete APV optimization for representative $(N,K)$ configurations. Both methods use the same initialization. The proposed method solves each projection with the Dykstra procedure, whereas the CVX baseline solves the same projection subproblem using CVX. Each entry is averaged over fifty repeated trials. It can be observed that the proposed method is substantially faster than the CVX baseline, showing the effectiveness of the proposed design.

\begin{figure}[t]
\centering
\subfigure[Convergence behavior of different total bandwidth.]{%
\includegraphics[width=0.47\columnwidth]{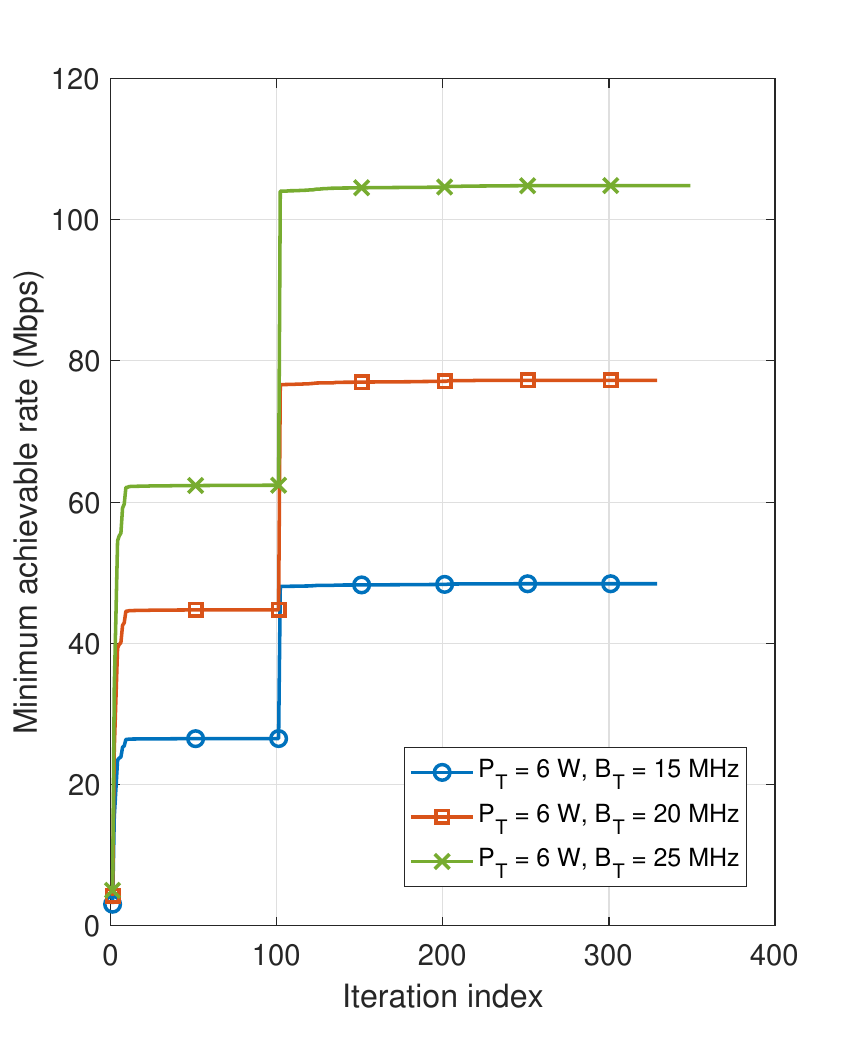}}
\hfill
\subfigure[Convergence behavior of different total power.]{%
\includegraphics[width=0.47\columnwidth]{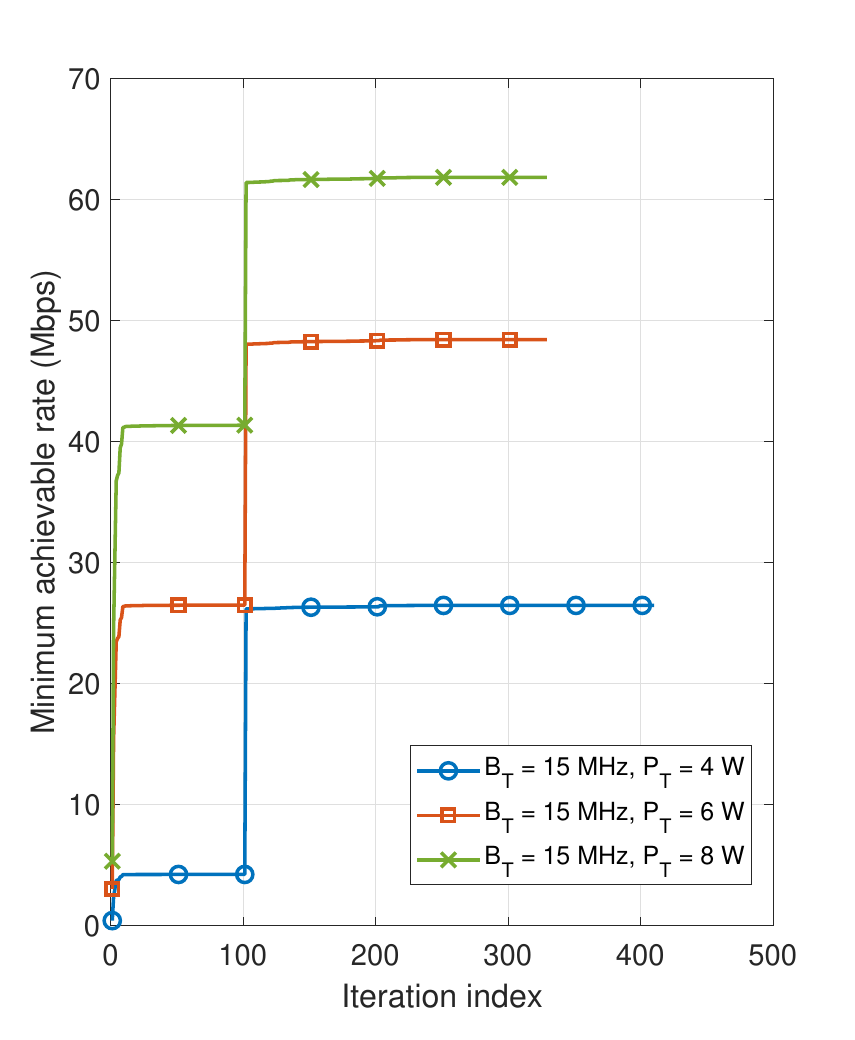}}
\caption{Convergence of the FDD-based joint optimization algorithm.}
\label{fig:conv}
\end{figure}

Fig.~\ref{fig:conv} shows the convergence of the proposed optimization algorithm for the FDD-based MA-aided ISAC system under the same target CRB threshold. The iteration index records one KKT bandwidth/power allocation update at the start of each outer iteration and the subsequent $T_{\rm in} = 100$ APV updates. Hence, the horizontal axis in Fig.~\ref{fig:conv} denotes the overall iteration index. For visual clarity, markers are displayed only at selected iteration indices. These sparse markers do not represent the actual number of executed iterations. In addition, the jump near index $100$ corresponds to bandwidth/power reoptimization based on the updated APV. This is because the sensing constraint is active in the first $(\eta_B,\eta_P)$ update, while the subsequent APV $\bm{x}$ updates make the sensing constraint looser. The second outer iteration then produces a pronounced increase in both $\eta_B$ and $\eta_P$, leading to more resources available for enhancing communication performance while meeting the sensing constraint.

\ifcommapvbenchmark
\begin{figure}[t]
\centering
\includegraphics[width=0.70\columnwidth, height = 1.9in]{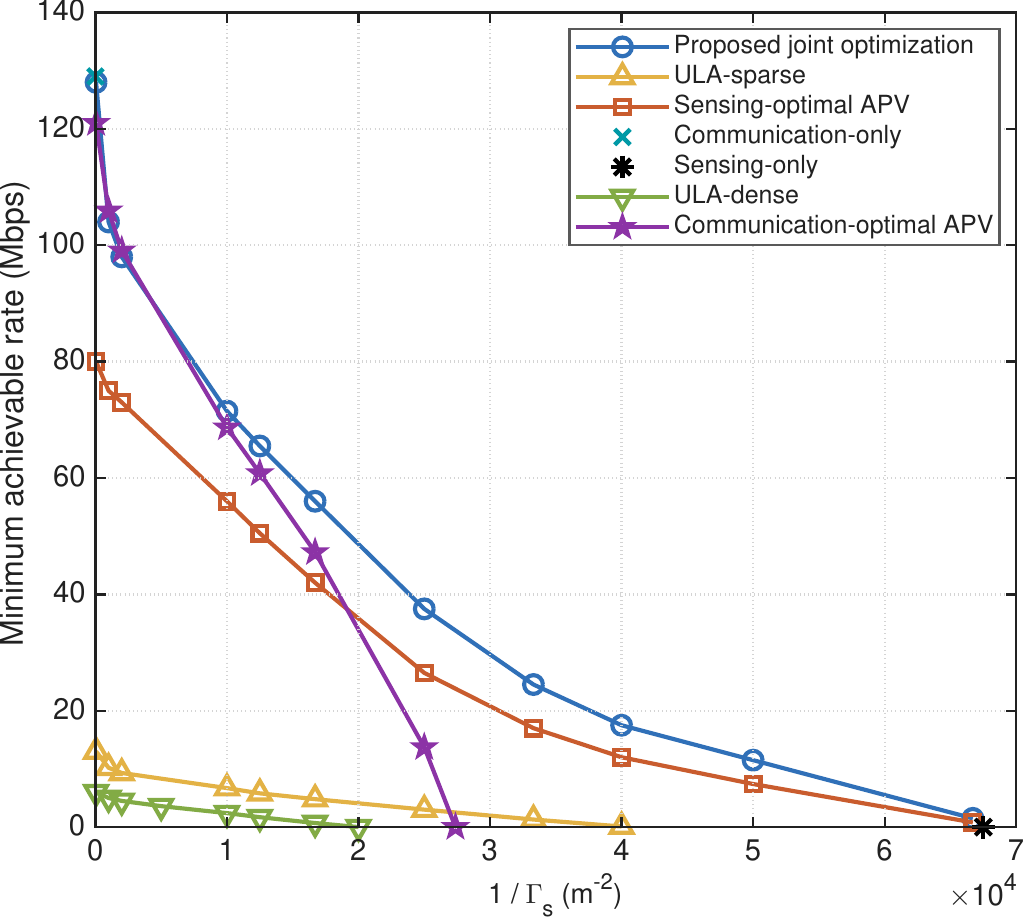}
\caption{Comparison of the minimum achievable rate versus the reciprocal of the CRB threshold for different schemes in the FDD scheme.}
\label{fig:CRB-rate}
\end{figure}
\else
\begin{figure}[t]
\centering
\includegraphics[width=0.70\columnwidth, height = 1.9in]{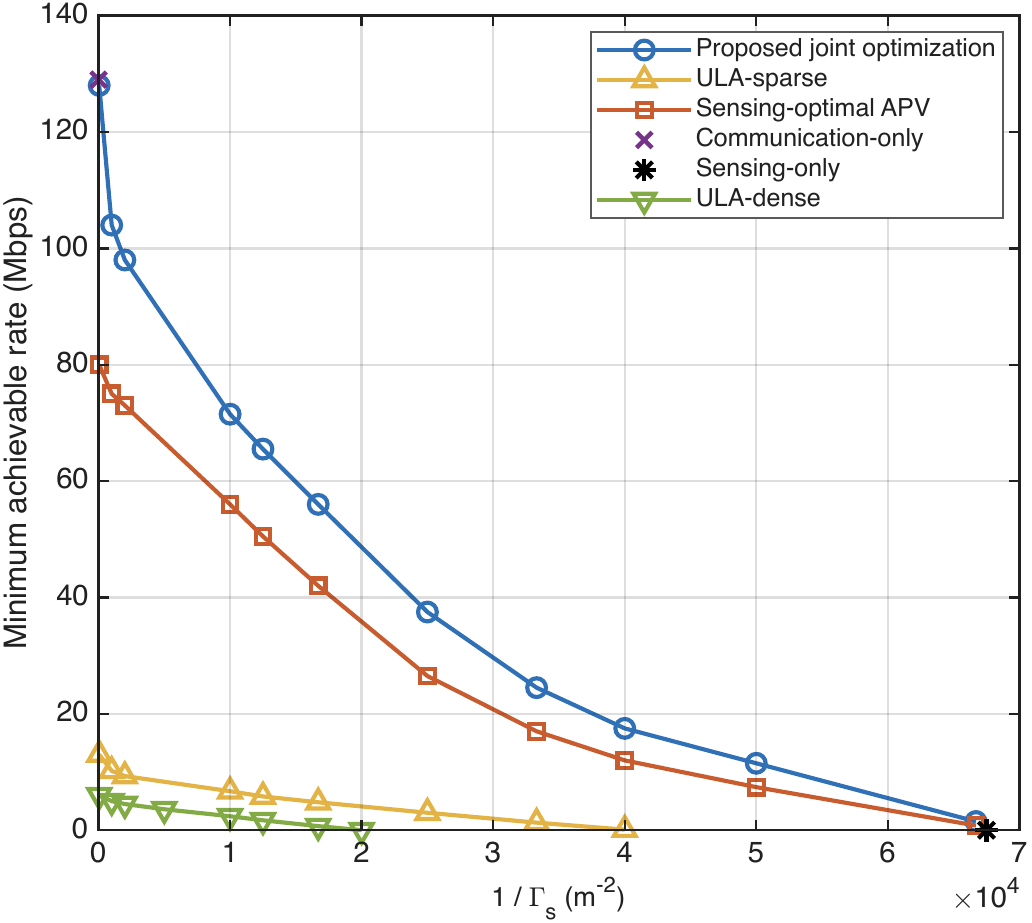}
\caption{Comparison of the minimum achievable rate versus the reciprocal of the CRB threshold for different schemes in the FDD scheme.}
\label{fig:CRB-rate}
\vspace{-0.3cm}
\end{figure}
\fi
Fig.~\ref{fig:CRB-rate} illustrates the communication-sensing tradeoff region characterized by the minimum achievable rate versus the reciprocal of the sensing CRB threshold. For the same sensing requirement where different sensing-constrained benchmarks are feasible, the proposed joint optimization achieves a higher minimum achievable rate than the FPA-based schemes. Moreover, the proposed design supports a wider tradeoff region, since it remains feasible under more stringent sensing requirements. As the sensing requirement becomes more stringent, the minimum achievable rates of all sensing-constrained schemes decrease and eventually approach zero, since bandwidth, transmit power, and spatial DoF have to be devoted to sensing.
The ULA-dense scheme has the smallest feasible reciprocal-CRB and the lowest rate since it needs large sensing bandwidth and power demand to compensate for its small aperture.
The communication-optimal APV performs well under loose sensing requirements. As the sensing constraint tightens, its smaller position variance increases the sensing-resource demand, causing its rate to decrease more rapidly and eventually fall below that of the sensing-optimal APV.

\subsection{TDD-Based Performance}

\ifcommapvbenchmark
\begin{figure}[t]
\centering
\subfigure[Minimum achievable rate under different sensing requirements.]{%
\includegraphics[width=0.47\columnwidth]{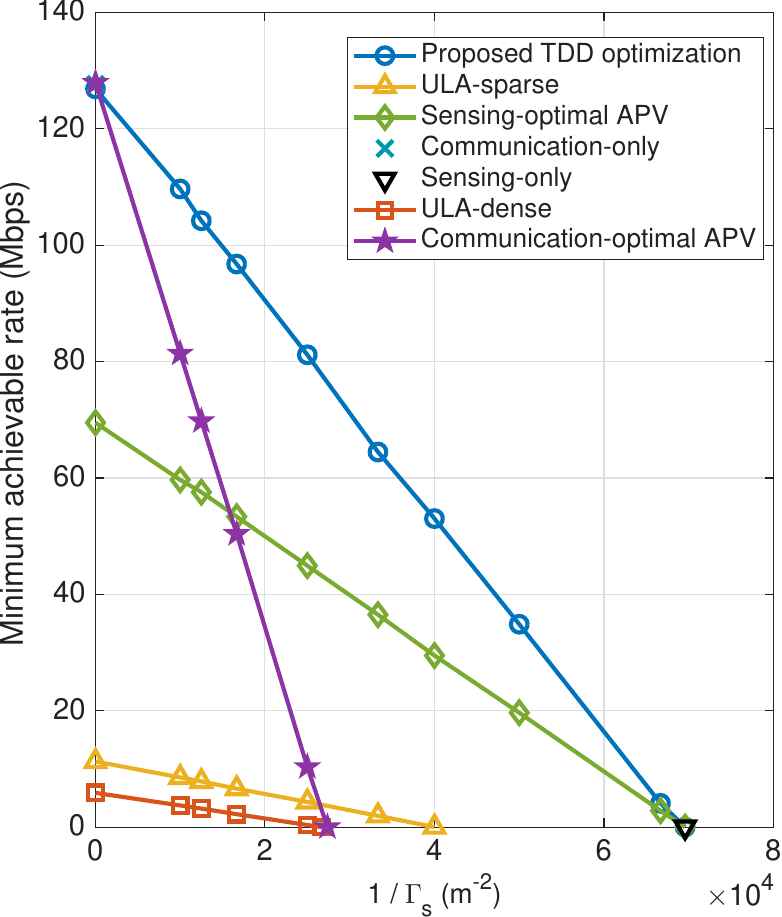}}
\hfill
\subfigure[Optimized sensing pulse number under different sensing requirements.]{%
\includegraphics[width=0.47\columnwidth]{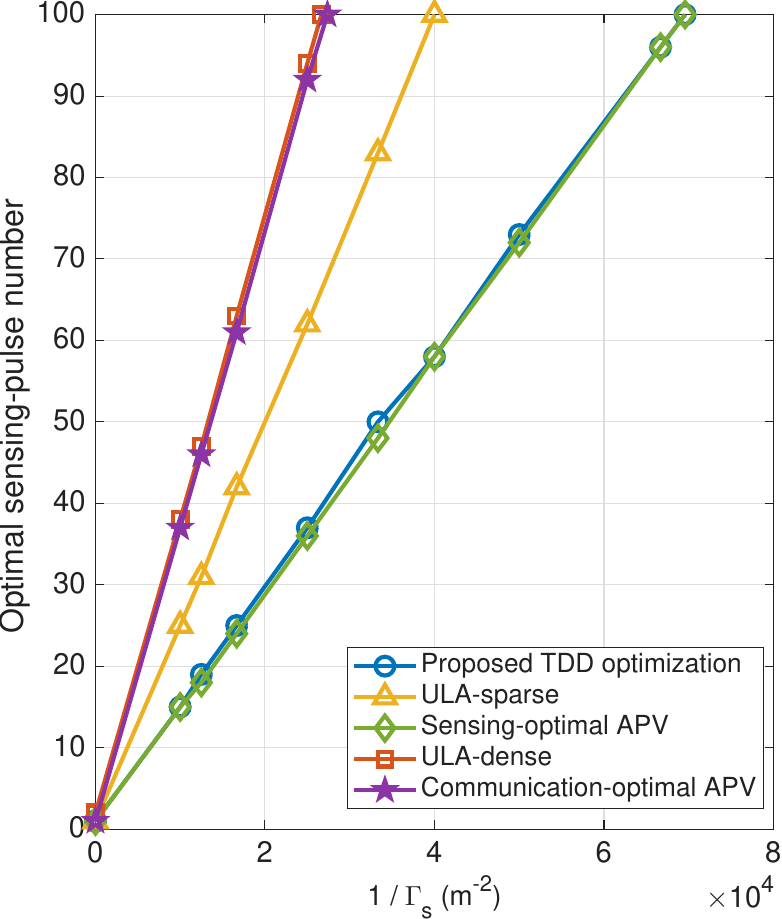}}
\vspace{0.3cm}
\caption{Communication-sensing tradeoff and the optimized sensing pulse number under different schemes in the TDD scheme.}
\label{fig:tdd-crb-tradeoff}
\end{figure}
\else
\begin{figure}[t]
\centering
\subfigure[Minimum achievable rate under different sensing requirements.]{%
\includegraphics[width=0.47\columnwidth]{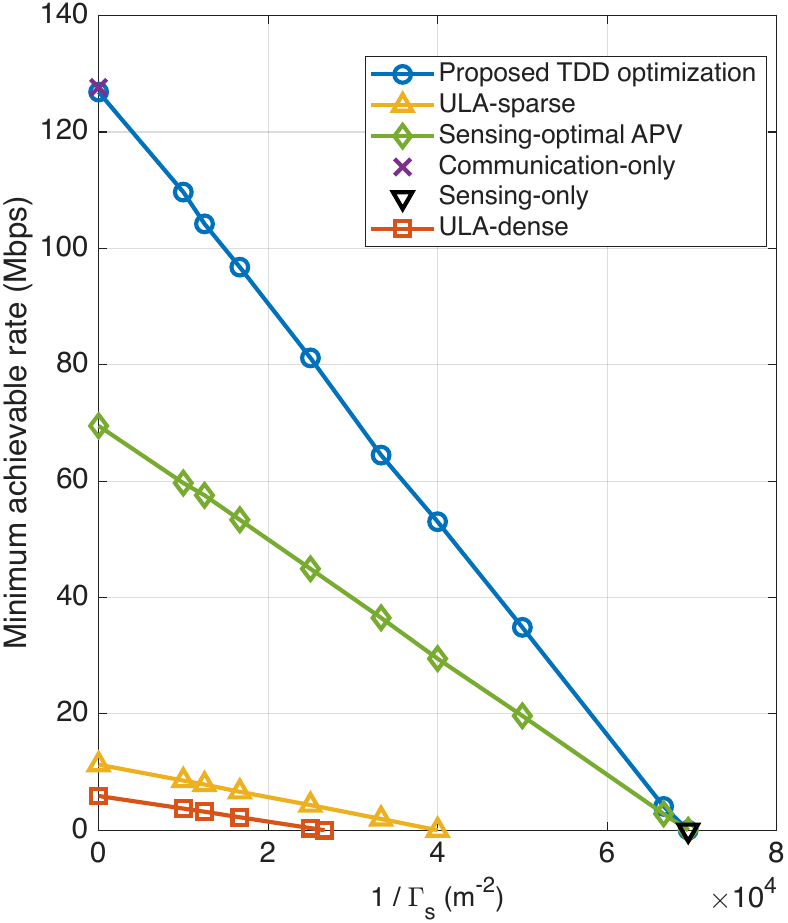}}
\hfill
\subfigure[Optimized sensing pulse number under different sensing requirements.]{%
\includegraphics[width=0.47\columnwidth]{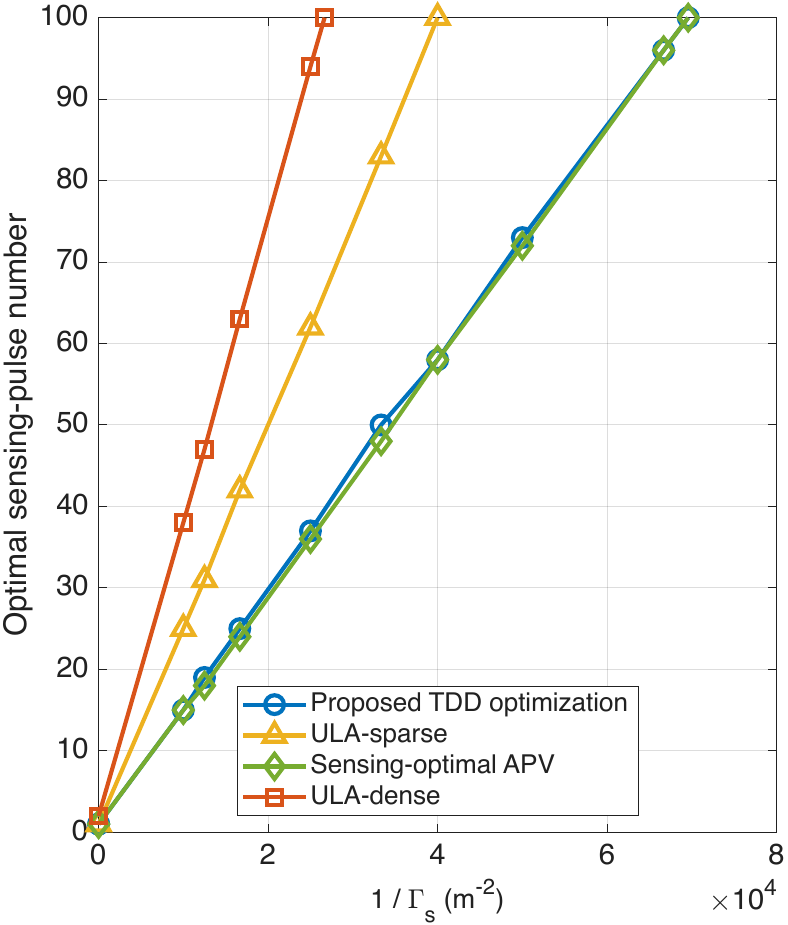}}
\vspace{-0.3cm}
\caption{Communication-sensing tradeoff and the optimized sensing pulse number under different schemes in the TDD scheme.}
\label{fig:tdd-crb-tradeoff}
\end{figure}
\fi

We next evaluate the performance of the TDD-based formulation.  
Fig.~\ref{fig:tdd-crb-tradeoff} illustrates the sensing-communication tradeoff under different sensing CRB requirements, where a larger $1/\Gamma_s$ represents a more stringent sensing requirement. As shown in Fig.~\ref{fig:tdd-crb-tradeoff}(a), tightening the sensing constraint leaves fewer time slots for communication and therefore reduces the minimum achievable rate. The proposed design maintains the same sensing performance as the sensing-optimal APV while achieving a substantially higher communication rate. In addition, the communication-only and sensing-only points represent the two limiting operating modes of the TDD scheme. 
Fig.~\ref{fig:tdd-crb-tradeoff}(b) shows the corresponding optimal number of sensing pulses for the benchmarks in Fig.~\ref{fig:tdd-crb-tradeoff}(a). The proposed design requires nearly the same number of pulses as the sensing-optimal APV, indicating a negligible sensing-performance loss. This result demonstrates the importance of jointly optimizing the APV and resource allocation to preserve sensing performance while improving multiuser channel orthogonality, which shows the effectiveness of our proposed design.

\ifcommapvbenchmark
\begin{figure}[t]
\centering
\subfigure[Minimum achievable rate versus the total time-slot budget.]{%
\includegraphics[width=0.47\columnwidth]{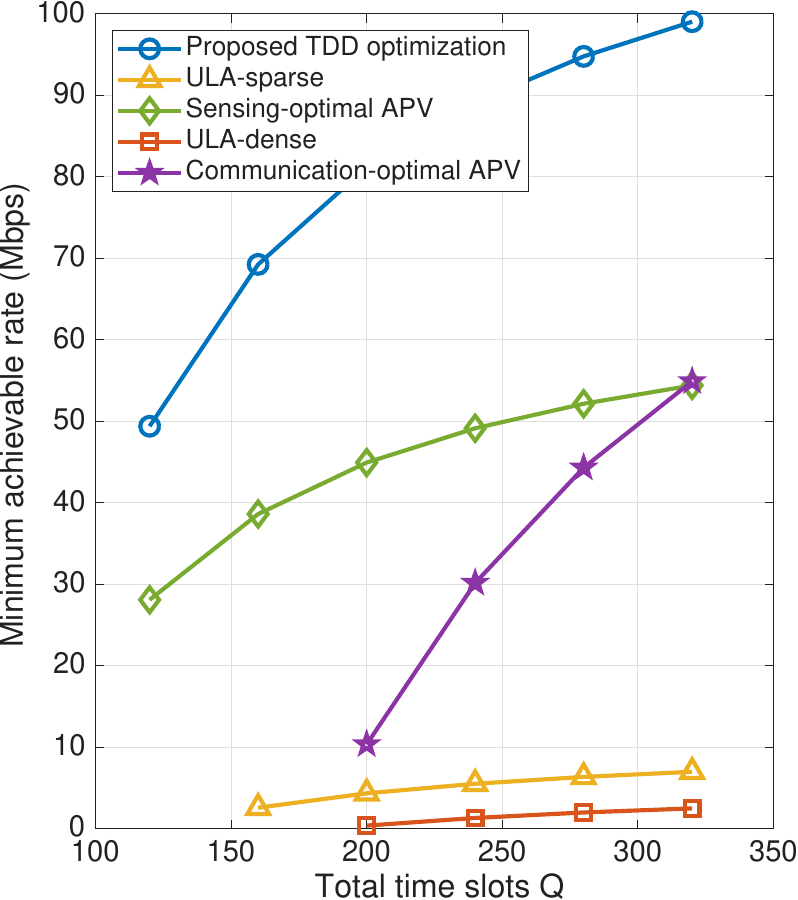}}
\hfill
\subfigure[Sensing-time fraction versus the total time-slot budget.]{%
\includegraphics[width=0.47\columnwidth]{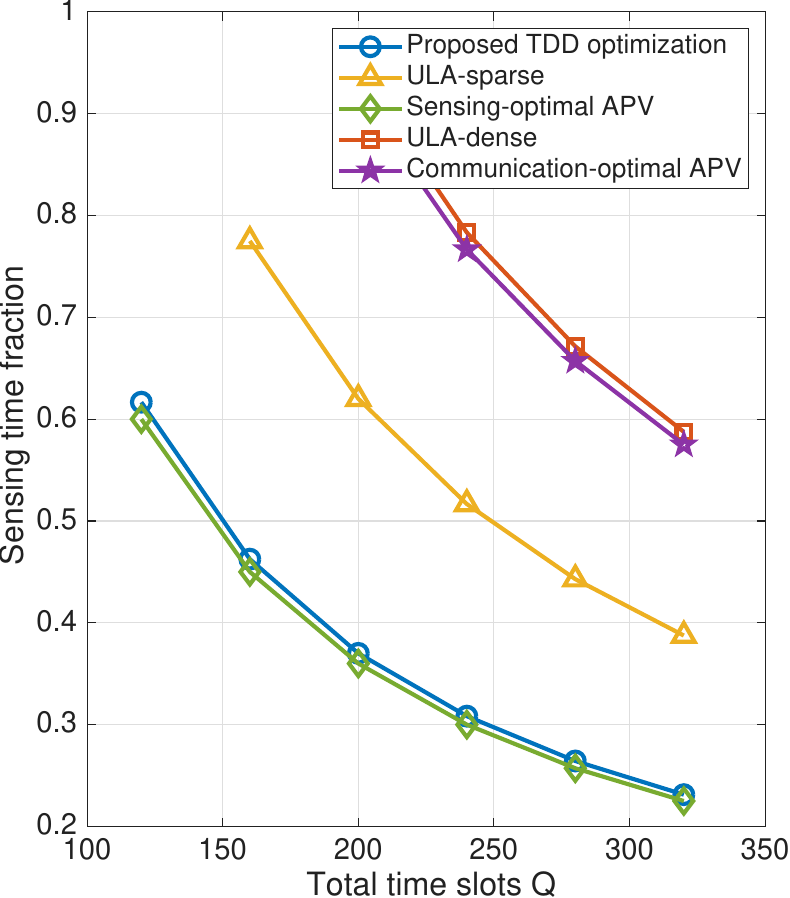}}
\vspace{0.3cm}
\caption{TDD performance under different total time-slot budgets.}
\label{fig:tdd-slot-budget}
\end{figure}
\else
\begin{figure}[t]
\centering
\subfigure[Minimum achievable rate versus the total time-slot budget.]{%
\includegraphics[width=0.47\columnwidth]{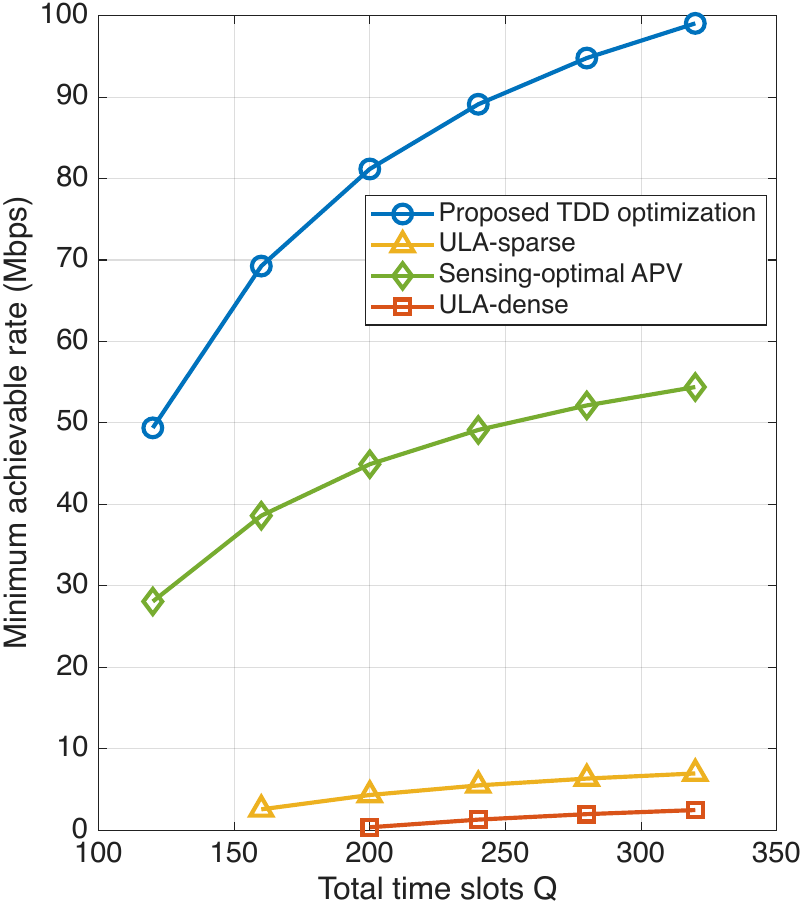}}
\hfill
\subfigure[Sensing-time fraction versus the total time-slot budget.]{%
\includegraphics[width=0.47\columnwidth]{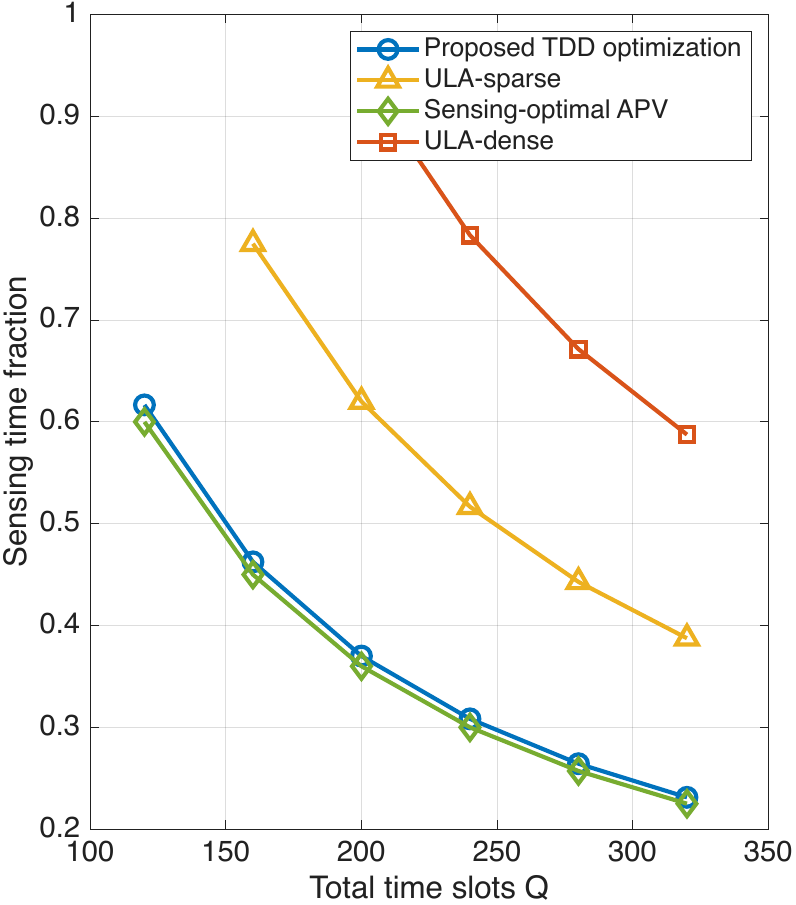}}
\vspace{0.3cm}
\caption{TDD performance under different total time-slot budgets.}
\label{fig:tdd-slot-budget}
\end{figure}
\fi

\begin{figure}[t]
\centering
\subfigure[$B_T=10$ MHz, $\Gamma_s=4\times10^{-4}$.]{%
\includegraphics[width=0.47\columnwidth]{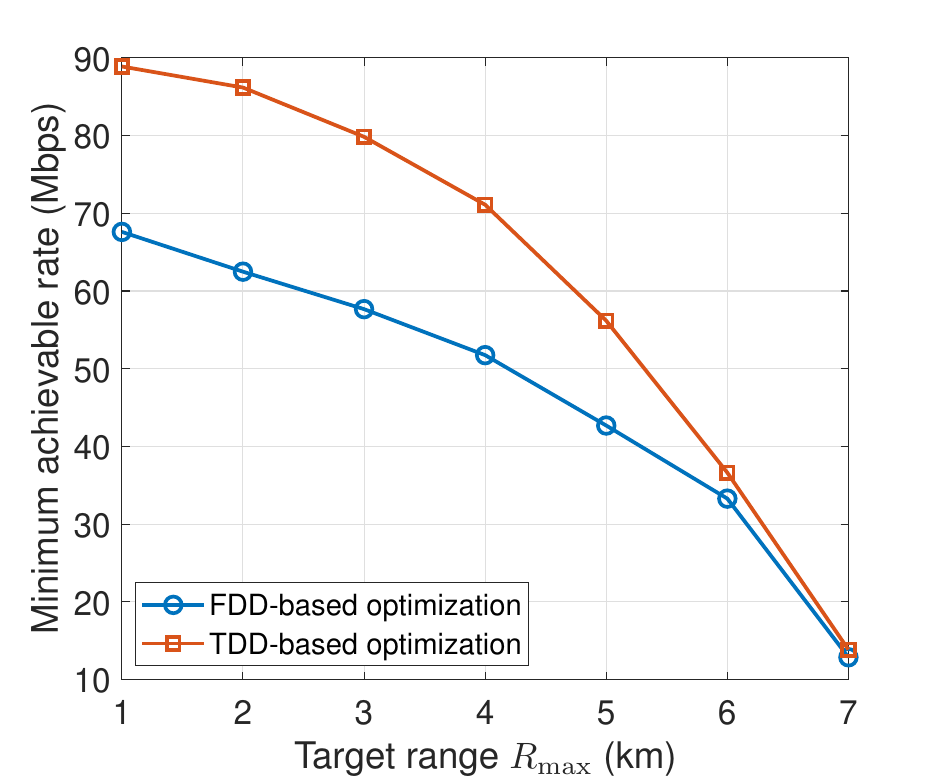}}
\hfill
\subfigure[$B_T=20$ MHz, $\Gamma_s=4\times10^{-4}$.]{%
\includegraphics[width=0.47\columnwidth]{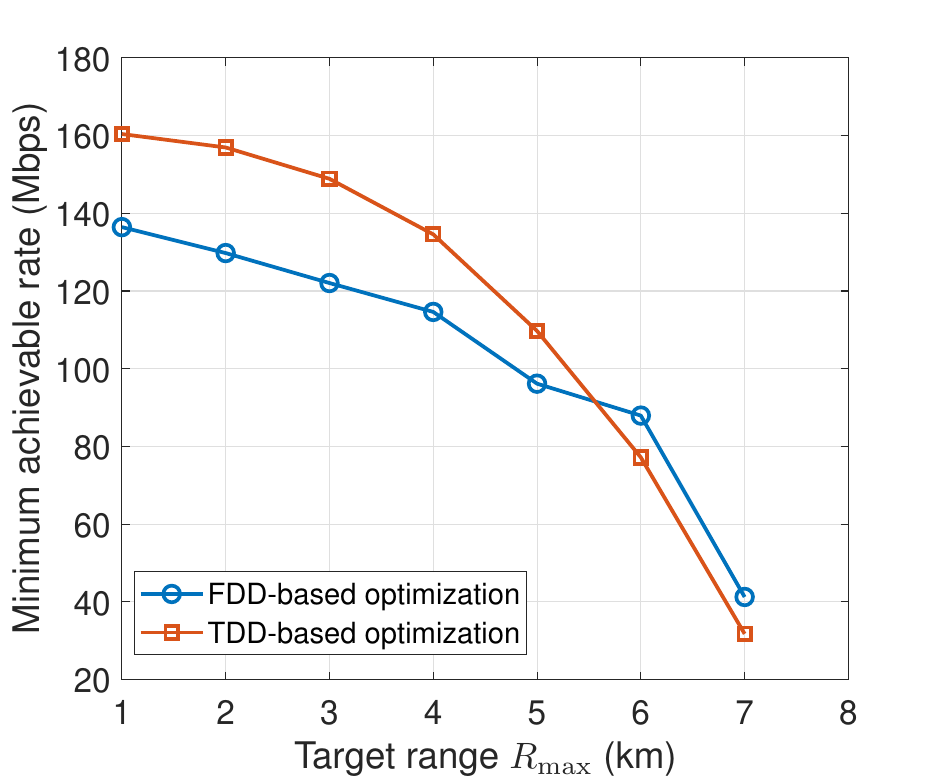}}
\\[-3mm]
\subfigure[$B_T=10$ MHz, $\Gamma_s=4\times10^{-3}$.]{%
\includegraphics[width=0.47\columnwidth]{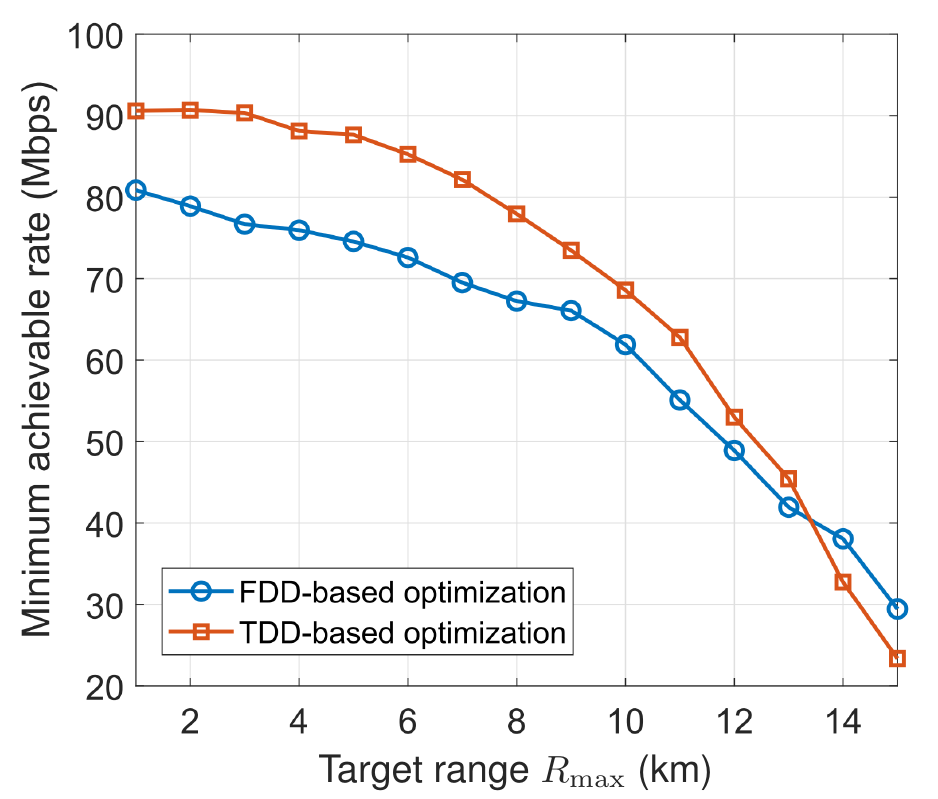}}
\hfill
\subfigure[$B_T=20$ MHz, $\Gamma_s=4\times10^{-3}$.]{%
\includegraphics[width=0.47\columnwidth]{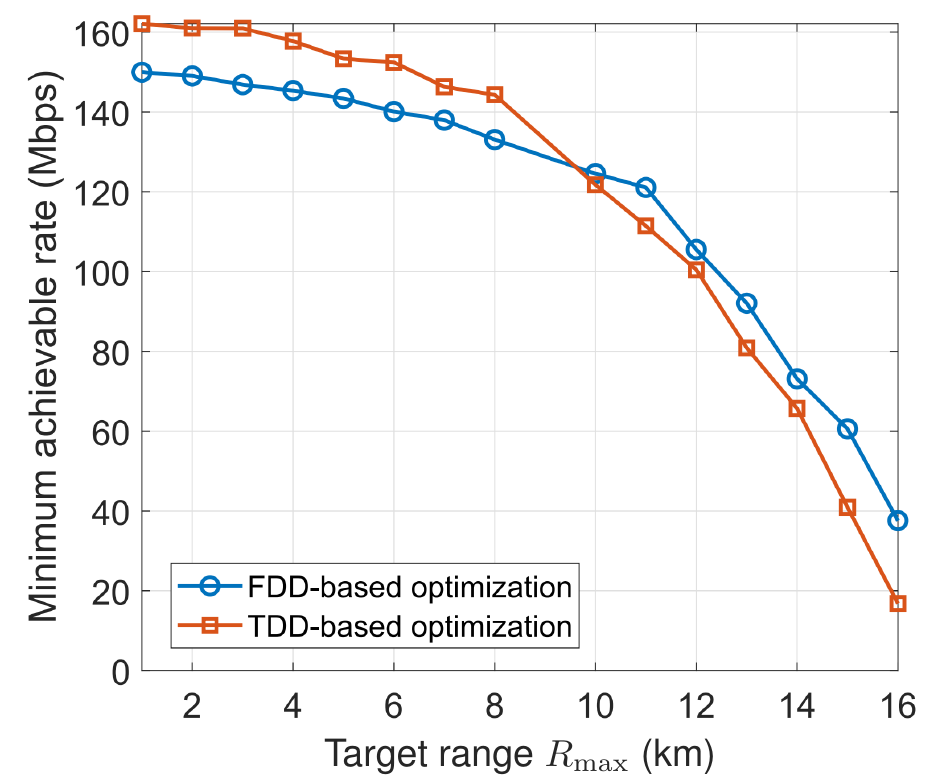}}
\caption{Minimum achievable rate versus target range for the proposed TDD- and FDD-based designs under different total bandwidths and sensing CRB thresholds.}
\label{fig:tdd-fdd-range-comparison}
\end{figure}

Fig.~\ref{fig:tdd-slot-budget} evaluates the impact of the total time-slot budget $Q$. As shown in Fig.~\ref{fig:tdd-slot-budget}(a), the minimum achievable rates of all schemes increase with $Q$, because the number of sensing pulses required to satisfy the prescribed CRB constraint remains approximately constant, and the additional time slots can therefore be primarily allocated to communication. This behavior is further supported by Fig.~\ref{fig:tdd-slot-budget}(b), where the sensing-time fraction decreases approximately inversely with $Q$. The ULA-sparse/dense schemes leave substantially fewer time slots for communication than the proposed design. Although the sensing-optimal APV saves several sensing pulses relative to the proposed design, it produces a much lower communication rate because it does not account for multiuser channel orthogonality. The communication-optimal APV requires a larger sensing-time fraction than the proposed design, which limits its achievable rate despite its favorable channel orthogonality. The proposed joint optimization therefore achieves the balance between sensing and communication performance.

\subsection{TDD/FDD Performance Comparison}

Fig. 8 compares the optimized TDD- and FDD-based rates versus the target range under different total bandwidths and sensing CRB thresholds. The rate crossover can be explained by both the sensing-load composition and the TDD sensing-time overhead. Since the angle-related load \(a\) increases with \(R_{\rm max}^2\), while the delay-related load \(d\) decreases with \(B_T^2\), increasing either \(R_{\rm max}\) or \(B_T\) increases the ratio \(\xi=a/d\) and shifts the sensing condition toward the angle-dominant regime. In addition, a larger \(R_{\rm max}\) requires a longer echo-reception interval and hence a longer sensing-pulse cycle, which further reduces the communication time available to TDD and makes FDD more favorable.
For \(\Gamma_s=4\times10^{-4}\) and \(B_T=10\) MHz, TDD maintains a higher rate over the evaluated target ranges. When \(B_T\) increases to \(20\) MHz, the reduced delay-related sensing load makes FDD more competitive, and FDD overtakes TDD at \(R_{\rm max}=6\) km. With the looser sensing requirement \(\Gamma_s=4\times10^{-3}\), both schemes remain feasible over a wider target-range interval, while increasing \(B_T\) again shifts the TDD/FDD crossover toward a shorter range.

\ifcommapvbenchmark
\begin{figure}[!t]
\centering
\subfigure[Achievable minimum communication rate, $\theta_k=\{\pm15^\circ,\pm50^\circ\}$]{%
\includegraphics[width=0.47\columnwidth]{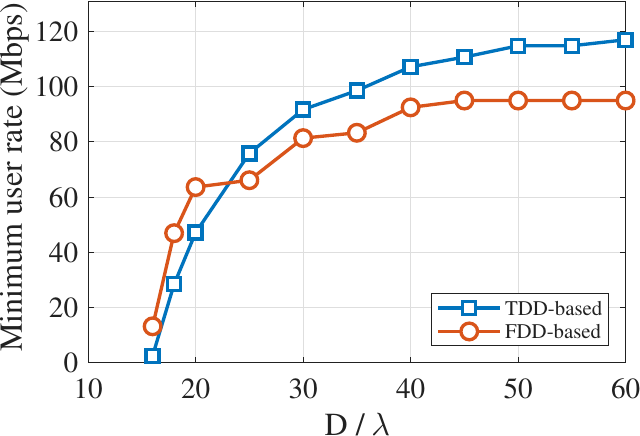}}
\hfill
\subfigure[Achievable minimum communication rate, $\theta_k=\{\pm1^\circ,\pm2^\circ\}$.]{%
\includegraphics[width=0.47\columnwidth]{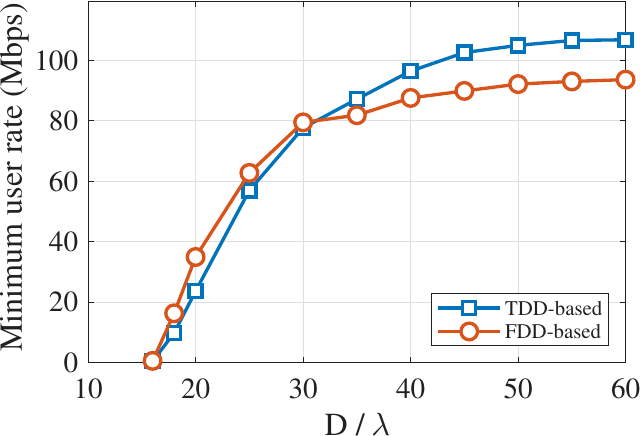}}
\\[-1mm]
\subfigure[Normalized SNR $q$, $\theta_k=\{\pm15^\circ,\pm50^\circ\}$.]{%
\includegraphics[width=0.47\columnwidth]{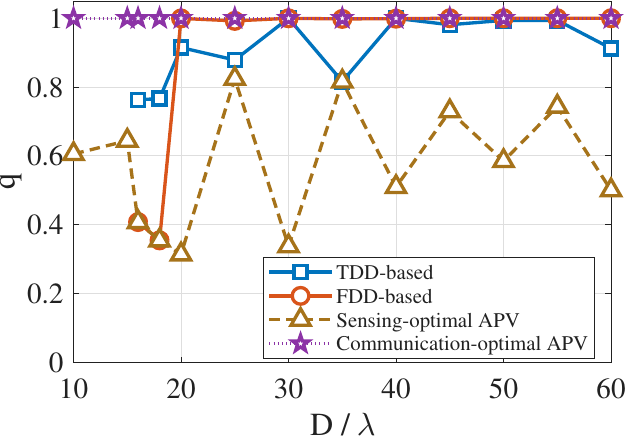}}
\hfill
\subfigure[Normalized SNR $q$, $\theta_k=\{\pm1^\circ,\pm2^\circ\}$.]{%
\includegraphics[width=0.47\columnwidth]{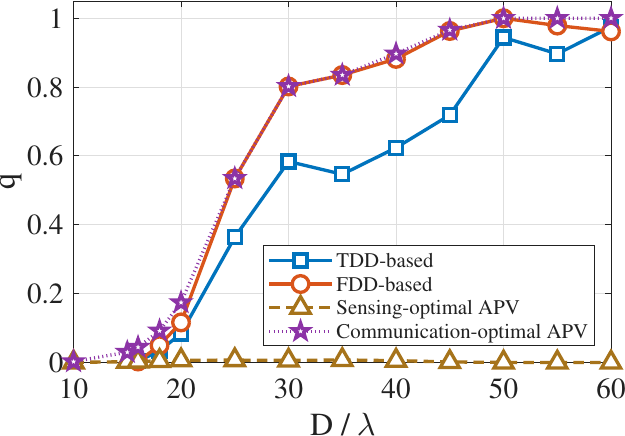}}
\\[-1mm]
\subfigure[Sensing load ratio $\xi$, $\theta_k=\{\pm15^\circ,\pm50^\circ\}$.]{%
\includegraphics[width=0.47\columnwidth]{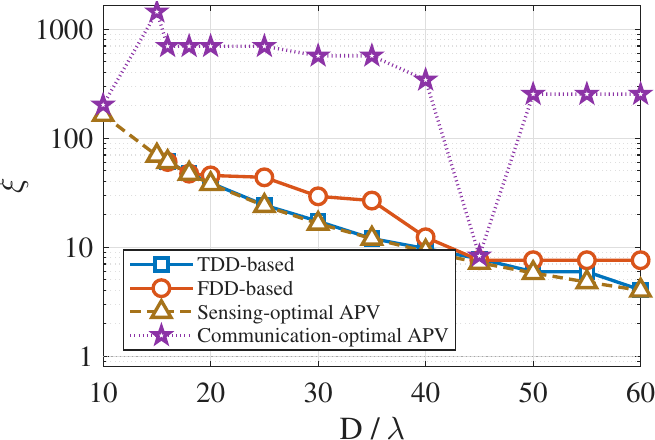}}
\hfill
\subfigure[Sensing load ratio $\xi$, $\theta_k=\{\pm1^\circ,\pm2^\circ\}$.]{%
\includegraphics[width=0.47\columnwidth]{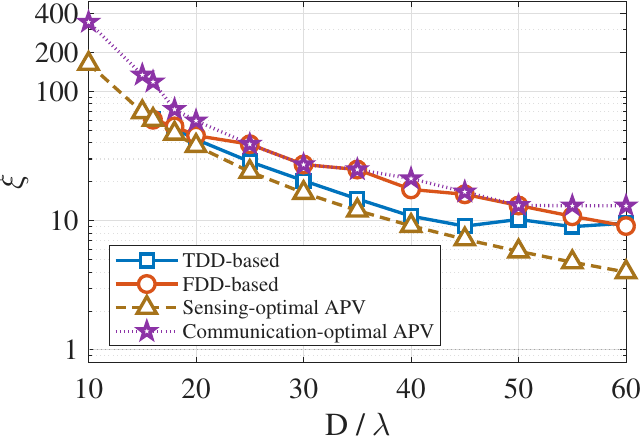}}
\\[-1mm]
\subfigure[Effective sensing-load ratio in FDD $\widetilde{\xi}_{\rm FDD}$, $\theta_k=\{\pm15^\circ,\pm50^\circ\}$.]{%
\includegraphics[width=0.47\columnwidth]{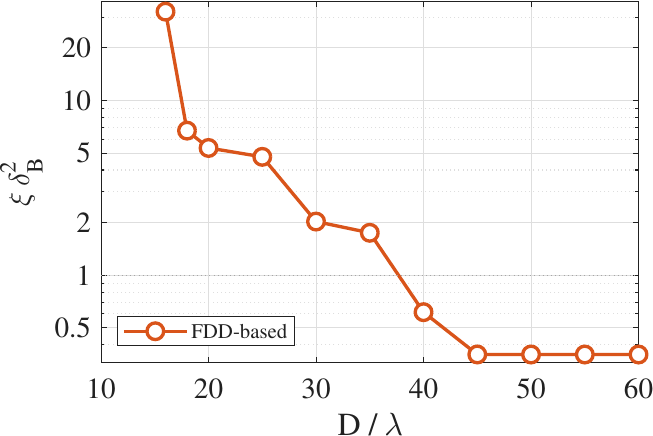}}
\hfill
\subfigure[Effective sensing-load ratio in FDD $\widetilde{\xi}_{\rm FDD}$, $\theta_k=\{\pm1^\circ,\pm2^\circ\}$.]{%
\includegraphics[width=0.47\columnwidth]{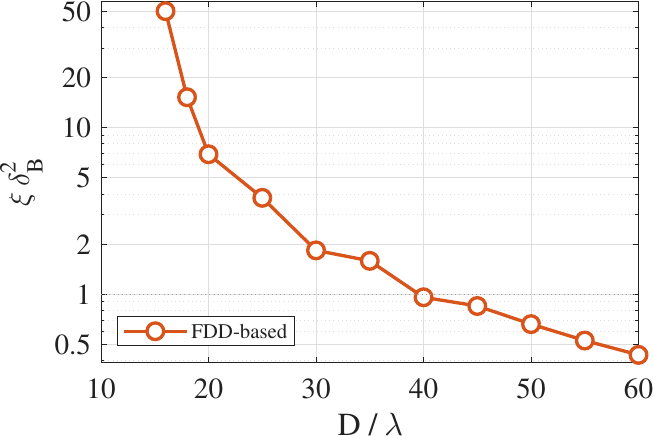}}
\\[-1mm]
\vspace{0.3cm}
\caption{Impact of MA movement region on TDD/FDD performance.}
\label{fig:four-user-combined}
\end{figure}
\else
\begin{figure}[!t]
\centering
\subfigure[Achievable minimum communication rate, $\theta_k=\{\pm15^\circ,\pm50^\circ\}$]{%
\includegraphics[width=0.47\columnwidth]{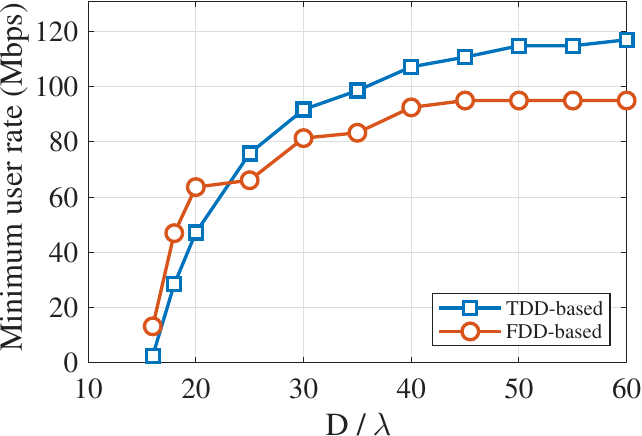}}
\hfill
\subfigure[Achievable minimum communicaiton rate, $\theta_k=\{\pm1^\circ,\pm2^\circ\}$.]{%
\includegraphics[width=0.47\columnwidth]{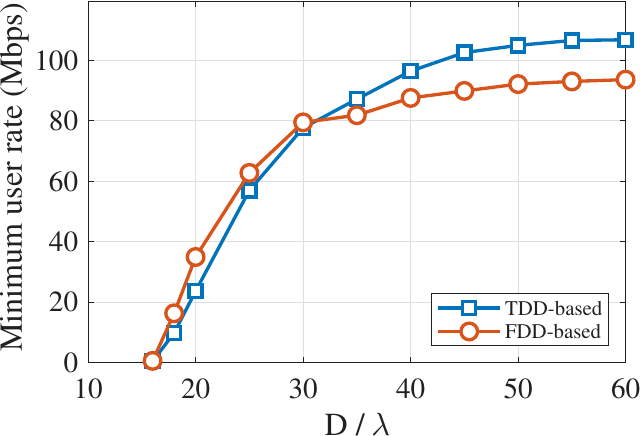}}
\\[-1mm]
\subfigure[Normalized SNR $q$, $\theta_k=\{\pm15^\circ,\pm50^\circ\}$.]{%
\includegraphics[width=0.47\columnwidth]{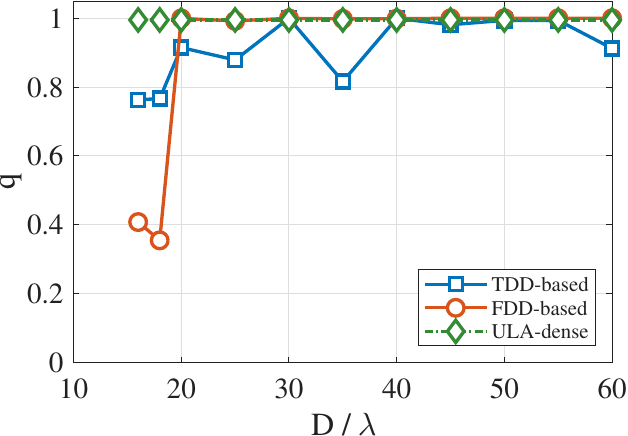}}
\hfill
\subfigure[Normalized SNR $q$, $\theta_k=\{\pm1^\circ,\pm2^\circ\}$.]{%
\includegraphics[width=0.47\columnwidth]{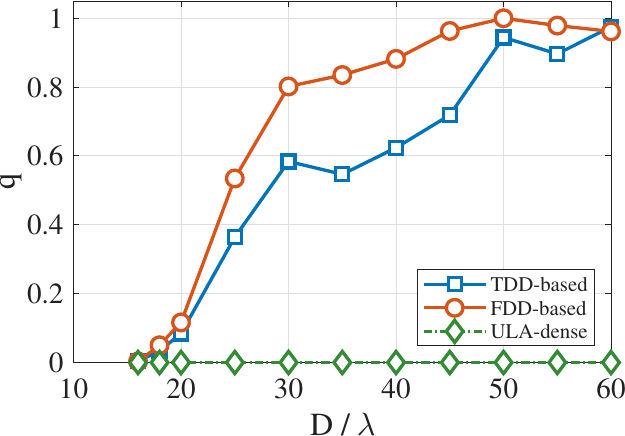}}
\\[-1mm]
\subfigure[Sensing load ratio $\xi$, $\theta_k=\{\pm15^\circ,\pm50^\circ\}$.]{%
\includegraphics[width=0.47\columnwidth]{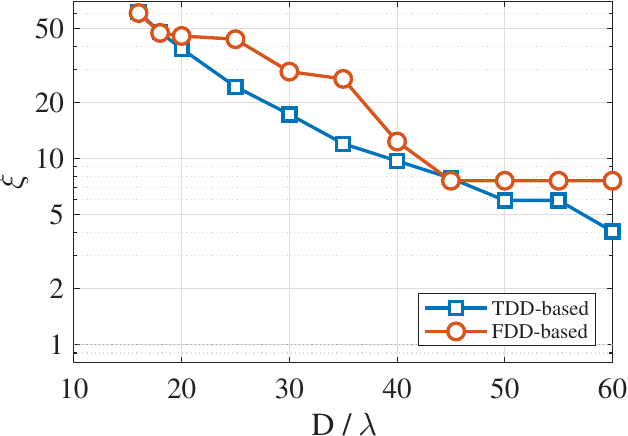}}
\hfill
\subfigure[Sensing-load ratio $\xi$, $\theta_k=\{\pm1^\circ,\pm2^\circ\}$.]{%
\includegraphics[width=0.47\columnwidth]{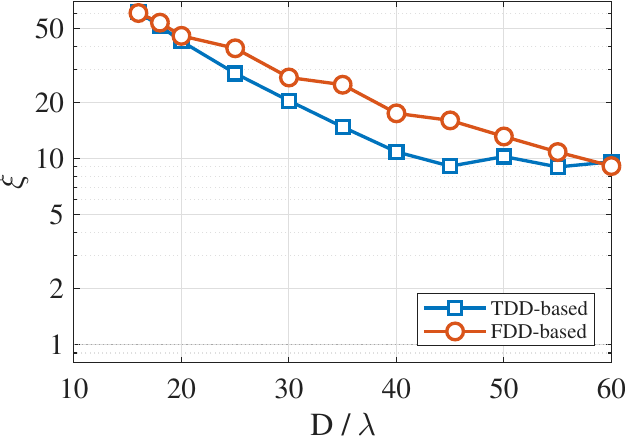}}
\\[-1mm]
\subfigure[Effective sensing-load ratio in FDD $\widetilde{\xi}_{\rm FDD}$, $\theta_k=\{\pm15^\circ,\pm50^\circ\}$.]{%
\includegraphics[width=0.47\columnwidth]{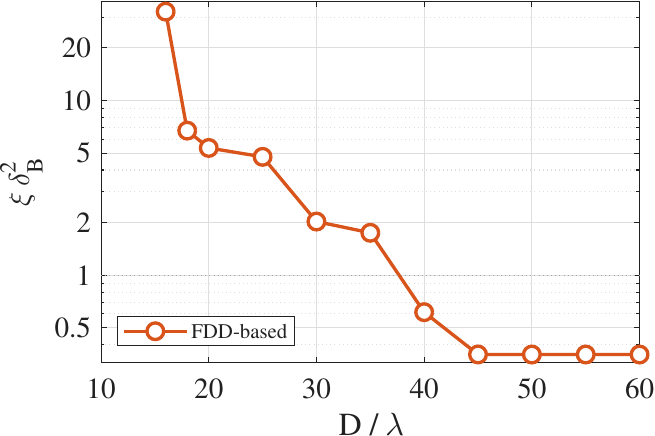}}
\hfill
\subfigure[Effective sensing-load ratio in FDD $\widetilde{\xi}_{\rm FDD}$, $\theta_k=\{\pm1^\circ,\pm2^\circ\}$.]{%
\includegraphics[width=0.47\columnwidth]{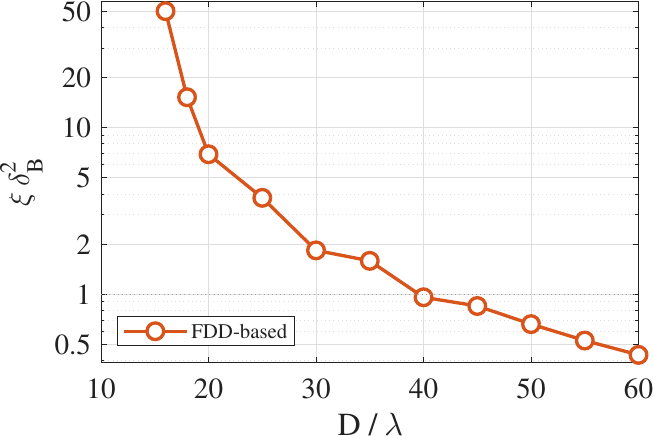}}
\\[-1mm]
\vspace{0.3cm}
\caption{Impact of MA movement region on TDD/FDD performance.}
\label{fig:four-user-combined}
\end{figure}
\fi

To further verify the roles of MA analyzed by the representative scenarios in Sec. V, Fig. 9 jointly examines the achievable rate, the normalized SNR $q$, sensing load ratio $\xi$, and effective FDD sensing-load ratio $\widetilde{\xi}_{\rm FDD}$ achieved by the proposed joint design as the MA movement region varies. 

For the relatively well-separated user angles shown in the left column of Fig. 9, the normalized SNR $q$ rapidly approaches one. This is consistent with Scenario~2 in Sec.~V, where favorable channel orthogonality can be maintained over a broad range of APVs, leaving more APV freedom for enlarging the sensing aperture. The fixed-APV references in Figs.~9(c) and 9(e) further show that APVs with similarly high $q$ can exhibit substantially different sensing-load ratios, while the sensing-optimal APV provides a reference for the smallest $\xi$.
The non-smooth variations in $q$ and $\xi$ is owing to the non-monotonic dependence of channel correlation on the APV and possible switching among APV configurations with similar communication performance but different antenna position variances.
For small movement regions, the limited sensing aperture results in a large $\xi$, and the resulting angle-dominant sensing condition favors FDD. As $D/\lambda$ increases, the achievable antenna position variance increases and the angle-related load decreases. Meanwhile, FDD bandwidth partitioning amplifies the delay-related CRB component, causing $\widetilde{\xi}_{\rm FDD}=\xi\delta_B^2$ to decrease faster than $\xi$. Together, these effects weaken FDD's advantage and allow TDD to eventually achieve a higher rate.

For the angularly clustered users in Figs.~9(b), 9(d), 9(f), and 9(h), the small $q$ and large $\xi$ correspond to Scenario~4 in Sec.~V. MA therefore needs to simultaneously improve multiuser channel orthogonality and enlarge the sensing aperture. The differences between the communication- and sensing-optimal APV references in Figs.~9(d) and 9(f) further highlight this tradeoff, showing that improving channel orthogonality and enlarging the sensing aperture can favor different APV configurations. As $D/\lambda$ increases, $q$ increases substantially while $\xi$ decreases. Different from the well-separated case, part of the MA optimization freedom is used to improve communication channel orthogonality. Consequently, FDD remains competitive over a larger movement region. As the movement region further increases, the reduced angle-related sensing load, together with the amplification of the delay-related CRB term caused by FDD bandwidth partitioning, decreases $\widetilde{\xi}_{\rm FDD}$, eventually allowing TDD to achieve a higher rate.

The optimized APV configurations further exhibit different tendencies under the two schemes. At small $D/\lambda$, the limited antenna position variance makes the sensing constraint tight. In this regime, enlarging the sensing aperture is important for both schemes because it relaxes the sensing requirement and expands the feasible region. As $D/\lambda$ increases, the optimized APV configurations reflect different tradeoffs between channel orthogonality and sensing aperture. In TDD, full-bandwidth sensing avoids the amplification of the delay-related CRB component caused by bandwidth partitioning. Increasing the antenna position variance therefore effectively reduces the angle-related component, which can reduce the required sensing time and leave more time for communication. Accordingly, TDD tends to achieve a larger antenna position variance and hence a smaller $\xi$, as shown in Figs.~9(e) and 9(f). In contrast, in FDD, bandwidth partitioning increases the relative importance of the delay-related CRB component, which cannot be reduced by aperture enlargement. As this component becomes increasingly important, the sensing benefit of further increasing the antenna position variance diminishes, making it relatively more beneficial to use the APV DoF to improve multiuser channel orthogonality. This helps explain why FDD tends to maintain a larger $q$, as shown in Figs.~9(c) and 9(d).

\section{Conclusion}
This paper investigated an MA-aided ISAC system with \textcolor{black}{the} TDD- \textcolor{black}{or} FDD-based \textcolor{black}{scheme}. A wideband target-position CRB was derived under a pulse-radar setup with pulse compression, which jointly captures the delay-induced sensing error bound determined by the sensing bandwidth and power and the angle-induced sensing error bound governed by the MA array geometry and sensing power. Based on this metric, a robust sensing constraint over a target uncertainty region was established, and joint APV and scheme-specific resource allocation problems were formulated to maximize the minimum achievable communication rate. To solve the resulting non-convex problems, an alternating optimization algorithm was developed for the FDD formulation, while the TDD formulation was addressed through a nested search over the sensing pulse number combined with APV optimization. 
Furthermore, the TDD/FDD comparison clarified the roles of MA by analyzing representative regimes characterized by multiuser channel orthogonality and the angle-to-delay sensing-load ratio. Numerical results demonstrated that APV optimization enlarges the communication-sensing tradeoff compared with FPA benchmarks. The TDD/FDD comparison further showed that increasing the target range or total bandwidth can \textcolor{black}{shift} the system toward an FDD-favorable regime. It further showed that FDD tends to maintain better multiuser channel orthogonality, while TDD tends to favor a larger antenna position variance and hence a larger effective sensing aperture.
\appendix

\subsection*{A. Proof of Proposition 1}

The target reflection coefficient $|\alpha|^2$ generally varies with the target range and scattering characteristics. In the considered pulse-radar model, however, the range information is obtained from delay estimation through $\tau=2R/c$, rather than from the received echo strength. Therefore, for a localized target uncertainty region $\mathcal C$, $|\alpha|^2$ is approximated by a representative constant reflection power to obtain a tractable robust sensing constraint. Under this approximation, \eqref{eq:CRB-pos} shows that the delay-induced part is independent of the target location, whereas the angle-induced part changes with the target location through the geometric factor $R^2/\cos^2\vartheta$.

Moreover, since $LT_{\rm slot}=T_{\rm PRI}=T_p+T_g$ and $T_g\ge\tau_{\max}\ge\tau$ for any $(R,\vartheta)\in\mathcal C$, each scheduled sensing-pulse cycle provides enough time for transmitting one pulse and receiving its echo for all target locations in $\mathcal C$. Hence, $L_{s,\mathcal C}$ is the number of valid sensing pulses within $T_s$. Define the worst-case geometric factor over $\mathcal C$ as
\begin{equation}
\Psi_{\mathcal C}
\triangleq
\max_{(R,\vartheta)\in\mathcal C}
\frac{R^2}{\cos^2\vartheta}
\le
\frac{R_{\max}^2}{\upsilon},
\label{eq:appendix-prop1-final}
\end{equation}
where the inequality follows from $R\le R_{\max}$ and $\cos^2\vartheta\ge\upsilon$ for all $(R,\vartheta)\in\mathcal C$. Substituting this upper bound into \eqref{eq:CRB-pos} yields \eqref{eq:CRB-constraint-general}. \textcolor{black}{Proposition 1 is thus proved.}

\vspace{-0.2cm}
\subsection*{B. Proof of Proposition 2}

To derive the gradient of $\mathcal J(\bm{x})$, we first utilize the matrix differential rules: $\mathrm{dtr}(\bm{S}^{-1})=-\mathrm{tr}(\bm{S}^{-2}\mathrm{d}\bm{S})$ and $\mathrm{d}\bm{S}=(\mathrm{d}\bm{H})^H\bm{H}+\bm{H}^H\mathrm{d}\bm{H}$ to express the differential of $\mathcal J(\bm{x})$ as
\begin{equation}
\mathrm{d}\mathcal J=-2\mathrm{Re}\left\{\mathrm{tr}(\bm{S}^{-2}\bm{H}^H\mathrm{d}\bm{H})\right\}.
\label{eq:appendix-gradient-diff-final}
\end{equation}
Utilizing the chain rule $\mathrm{d}\bm{H}=\sum_{n=1}^N\frac{\partial\bm{H}}{\partial x_n}\mathrm{d}x_n$, the $n$-th entry of the gradient $\nabla \mathcal J(\bm{x})$ is given by
\begin{equation}
[\nabla \mathcal J(\bm{x})]_n
=-2\mathrm{Re}\left\{\mathrm{tr}\left(\bm{S}^{-2}\bm{H}^H\frac{\partial\bm{H}}{\partial x_n}\right)\right\}.
\label{eq:appendix-gradient-entry-final}
\end{equation}
Note that $\frac{\partial\bm{H}}{\partial x_n}$ is a sparse matrix where only the $n$-th row is non-zero, with $\frac{\partial H_{n,k}}{\partial x_n}=jk_\lambda\sin\theta_kH_{n,k}$. Therefore, the trace in \eqref{eq:appendix-gradient-entry-final} reduces to a summation over $k$:
\begin{equation}
[\nabla \mathcal J(\bm{x})]_n
=2k_\lambda\,\mathrm{Im}\left\{\sum_{k=1}^K\sin\theta_k[\overline{\bm{G}}]_{n,k}H_{n,k}\right\}.
\label{eq:appendix-gradient-sum-final}
\end{equation}
Finally, the gradient is compactly represented as
\begin{equation}
\nabla \mathcal J(\bm{x})
=2k_\lambda\,\mathrm{Im}\left((\overline{\bm{G}(\bm{x})}\odot\bm{H}(\bm{x}))\bm{c}\right),
\label{eq:appendix-gradient-final}
\end{equation}
where $\odot$ and $\overline{(\cdot)}$ denote the Hadamard product and complex conjugate, respectively. \textcolor{black}{Proposition 2 is thus proved.}

\end{document}